\documentclass[onecolumn]{article}    

\usepackage{graphicx}          
                               
\usepackage{amssymb}
\usepackage{amsthm}
\usepackage{mathtools}
\usepackage{authblk}
\usepackage{hyperref}

\usepackage{bm}
\usepackage{amsmath}
\usepackage{amsfonts}
\usepackage{amsmath}
\usepackage{amsthm}    
\usepackage{longtable} 
\usepackage{pdflscape} 
\newcommand*{\earth}{{\oplus}}
\usepackage{subfigure}
 \usepackage{hyperref} 

\usepackage{mathabx}
\usepackage{blindtext}

\usepackage{multirow}

\usepackage{mathtools}
\usepackage{siunitx}
\usepackage{authblk}
\usepackage{amsfonts}
\renewcommand\vec{\mathbf}
\usepackage{float}
\usepackage{pdflscape} 
\usepackage{subfigure}
 \usepackage{hyperref} 
\usepackage{blindtext}

\usepackage{multirow}

\usepackage[a4paper, rmargin = 2.5cm, lmargin = 2.5cm]{geometry}

\renewcommand\vec{\mathbf}

\usepackage{float}
\usepackage{epstopdf}
\AppendGraphicsExtensions{.tif}

\title{Shallow Encounters' Impact on Asteroid Deflection Prediction and Implications on Trajectory Design~\footnote{Manuscript accepted for publication by the AIAA's Journal of Guidance, Control, and Dynamics. DOI: soon}}

\author[a]{Rodolfo Batista Negri~\footnote{Postdoctoral Researcher, Institute of Science and Technology, Av. Cesare Monsueto Giulio Lattes 1201, São José dos Campos. Email: rodolfo.negri@unifesp.br}}
\author[b]{Antônio F. B. A. Prado~\footnote{Postgraduate Division, 12227-010, São José dos Campos, SP, Brazil.}}
\affil[a]{Federal University of São Paulo}
\affil[b]{National Institute for Space Research}

\begin{document}

\maketitle

 \begin{abstract}
Analytical approximations are commonly employed in the initial trajectory design phase of a mission to rapidly explore a broad design space. In the context of an asteroid deflection mission, accurately predicting deflection is crucial to determining the spacecraft's trajectory that will produce the desired outcome. However, the dynamics involved are intricate, and simplistic models may not fully capture the system's complexity. This study assesses the precision and limitations of analytical models in predicting deflection, comparing them to more accurate numerical simulations. The findings reveal that encounters with perturbing bodies, even at significant distances (a dozen times the radii of the sphere of influence of the perturbing planet), can markedly disturb the deflected asteroid's trajectory, resulting in notable disparities between analytical and numerical predictions. The underlying reasons for this phenomenon are explained, and provisional general guidelines are provided to assist mission analysts in addressing such occurrences. By comprehending the impact of shallow encounters on deflection, this study equips designers with the knowledge to make informed decisions throughout the trajectory planning process, enhancing the efficiency and effectiveness of asteroid deflection missions.
 \textbf{Keywords:} asteroid deflection; trajectory design; mission analysis; astrodynamics; shallow encounters.
\end{abstract}




\section{Introduction}
\label{sec:intro}

The kinetic impulse is considered the most effective technique for asteroid deflection at the current scientific and technological level \cite{sanchez2009multicriteria,weisbin2015comparative,thiry2017statistical,anthony2018asteroid,sanchez2020evaluation}, and as a result, it has been the focus of the majority of studies. For that same reason, it was recently tested by {NASA's (National Aeronautics and Space Administration)} DART (Double Asteroid Redirection Test) mission \cite{cheng2015}, which impacted the binary asteroid 65803 Didymos in late 2022. This mission demonstrated the impulsive kinetic impact strategy by aiming at the smaller body in the binary system, providing a real-scale experiment for the matrix $\boldsymbol{\beta}$~\cite{cheng2015}, which represents the magnifying effect in the transfer of momentum to the target due to ejection of debris from the impact. The transfer of momentum in the DART mission is being analyzed from ground-based observations and data from the Italian Space Agency's CubeSat {LICIACube (Light Italian CubeSat for Imaging of Asteroids)}, which arrived shortly after the impact, providing important images of the ejected material \cite{cheng2020dart,dotto2021liciacube}. Additionally, an important step was testing the autonomous GN\&C, the SMARTNAV system, which ensured the detection of the smaller body and guided the spacecraft for a hyper-velocity collision with the target~\cite{chen2018smartnav}. 

To make deflection missions possible, the prediction of deflection plays a crucial role in trajectory analysis and planning. As is the case in any deep space mission, trajectory design is crucial for finding a cost-effective solution within time constraints (which is even more important for a deflecting scenario). In a preliminary phase of the design process, the common approach is to resort to analytical approximations within the optimization routines to explore a larger design space promptly, which will indicate the most interesting regions for later refinement in more elaborate simulations~\cite{conway2010spacecraft}. In the context of asteroid deflection, the prediction of deflection is crucial for finding the trajectory that will allow the spacecraft to arrive for deflection in conditions that will produce the necessary deflection within safe margins.

The history of research on asteroid deflection includes various analytical estimates and optimization problems to find the minimum impulse required for deflection. Ahrens and Harris \cite{ahrens1992deflection} derived one of the earliest analytical estimates for the impulse needed to deflect an asteroid on a collision course with Earth. Their approach considered the two-body problem involving the asteroid and the Sun, applying a tangential impulse to the asteroid's velocity. Park and Ross \cite{park1999} formulated an optimization problem for asteroid deflection, focusing on a planar two-body problem and using Lagrange coefficients for time propagation. They found that the optimal impulse should be applied at the asteroid's perihelion, and, in longer interception times, it tends to align parallel to its orbital velocity. Conway \cite{conway2001} expanded this optimization problem to a three-dimensional case, considering an interception shortly before the predicted collision (within one year). Conway's results confirmed that for interceptions close in time to the predicted collision, the radial component of the impulse becomes more relevant, while the component normal to the asteroid's orbit is not optimal at all.

Ross et al. \cite{ross2001} introduced a third-body approximation, considering the gravitational effect of the Earth at the moment of collision. Accounting for Earth's gravitational attraction resulted in a significant increase in the optimal impulse, particularly for asteroids with semi-major axes close to 1 AU (probably due to lower relative velocities and stronger gravitational effects). Park and Mazanek \cite{park2003} synthesized all the previously mentioned approaches, considering Earth's gravitational attraction at the moment of collision, the three-dimensional case, and different interception times, confirming the general findings of previous studies.

Carusi et al. \cite{carusi2002} derived an analytical estimate based on Öpik's theory \cite{opik1976interplanetary}, considering the gravitational focusing effect of the Earth. Izzo \cite{izzo2005} developed a similar formulation based on the work of Scheeres and Schweickart \cite{scheeres2004mechanics}, but with more general equations that can be extended to low-thrust deflection strategies. Izzo et al. \cite{izzo2006optimal} and Izzo \cite{izzo2007optimization} further generalized and improved this work, incorporating the concept of the b-plane and resonant returns as introduced by Valsecchi et al. \cite{valsecchi2003resonant}, which allowed them to find optimal deflections that avoid the existence of gravitational keyholes in the b-plane.

The concept of a gravitational keyhole is significant as it advances in the consideration of a three-body problem for the deflection assessment. By using the theory of resonant encounters developed by Valsecchi et al. \cite{valsecchi2003resonant}, it is possible to calculate points in the b-plane in which the interception could result in a resonance return to a future collision with Earth. However, numerical computation of these points is challenging, leading Izzo et al. \cite{izzo2006optimal} and Izzo \cite{izzo2007optimization} to focus on calculating optimal impulses that avoid the existence of gravitational keyholes.

Vasile and Colombo \cite{vasile2008optimal} derived an analytical formulation using Gauss's planetary equations and studied strategies for optimal impulses. Their approach is more general than that of Izzo \cite{izzo2007optimization}, as it considers all effects on the mean anomaly, rather than just the perturbation in mean motion. {This development was further extended, accounting for low-thrust deflection with analytical and semi-analytical approaches~\cite{colombo2009semi,zuiani2012evidence}.} Although Vasile and Colombo \cite{vasile2008optimal} did not incorporate resonant encounters into their optimization, they performed a brief analysis using a circular restricted three-body problem, corroborating results obtained by Park and Mazanek \cite{park2003}.

Despite the progress made in studying deflection using two-body and three-body approximations, little attention has been given to the continuous gravitational interactions among three or more bodies, asteroid-Sun-Earth-other planets. It raises whether there are general relationships to be discovered in such scenarios. This gap in understanding motivates this work to explore the broader and more complex framework of the influence of other celestial bodies on the heliocentric trajectory of an asteroid between the moment of deflection and the predicted impact epoch. Indications of this behavior were initially observed in Negri et al. \cite{negri2019third}, which analyzed Jupiter's effects on asteroid deflection using the BCR4BP (bicircular restricted four-body problem)~\cite{negri2020generalizing}. Notably, these effects are not limited to Earth alone, and they can occur at distances well beyond the perturbing body's sphere of influence (i.e., not treated by the gravitational keyhole methods) as documented in Negri~ \cite{negri2022tese}. 

The primary scientific contributions of this work lie in revealing and explaining the relationship between distant shallow encounters of an asteroid with a planet and how these interactions influence asteroid deflection. This understanding holds significant importance for spacecraft trajectory design aimed at deflecting asteroids. Designing trajectories for interplanetary missions is inherently complex, with various dynamic interactions occurring throughout the trajectory, including the use of multiple gravity-assists.

The use of analytical models for predicting deflection and a comprehensive understanding of their limitations is of paramount importance in the trajectory design process. They enable trajectory designers to explore a broad design space confidently, knowing that the initial search results will remain reliable when later applied to more complex and accurate simulations. As a result of our research, we provide valuable guidelines for future trajectory designers tasked with considering asteroid deflection during the preliminary design phase of a mission. By gaining insights into how shallow encounters with planets can influence a deflection mission, our study equips designers with the essential knowledge to make informed decisions throughout the trajectory planning process. 

To achieve this goal, our study adopts an n-body problem that includes all the planets of the Solar System, employing the Negri and Prado \cite{negri2022nbody} approximation as the more accurate simulation environment for comparison and analysis. For impulsive deflection, we consider the analytical models proposed by Vasile and Colombo \cite{vasile2008optimal}, while for low-thrust deflections, we utilize the models introduced by Izzo \cite{izzo2007optimization}. 

\section{Fundamental Concepts}

This section introduces key concepts that will be utilized throughout this study, specifically the Gravitational Focusing discussed in Section \ref{sec:FocGrav} and the B-plane elaborated upon in Section \ref{sec:bplane}.

\subsection{Gravitational Focusing}
\label{sec:FocGrav}

\begin{figure}[!ht]
\centering
\includegraphics[width = .5\textwidth]{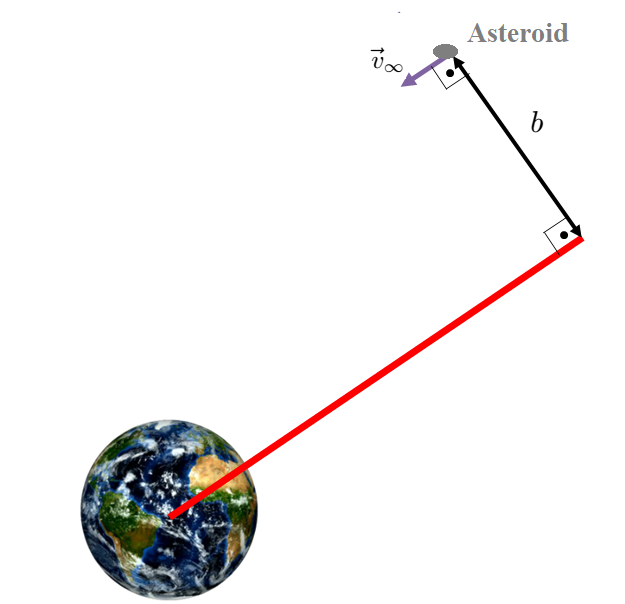}
\caption{Representation of the approach between the asteroid and Earth.}
\label{fig:approach}
\end{figure}

Gravitational focusing considers the gravitational interaction between two particles to predict a potential impact between them \cite{barnes2011}. Assuming that one of the particles is the asteroid and the other is the Earth, the conservation of energy and angular momentum in the two-body problem allows obtaining the impact parameter as:
\begin{equation}
\label{eqn:approach}
b = \sqrt{r_p^2 + \frac{\mu}{v_\infty^2} r_p}.
\end{equation}
where $b$ represents the impact parameter, which is perpendicular to the asteroid's arrival velocity, as represented in Fig. \ref{fig:approach}, it is worth noting that $b$ is equivalent to the semi-minor axis of a hyperbolic orbit. The variable $v_\infty$ is the geocentric velocity of the asteroid at infinity (approaching velocity), {$\mu$ is the gravitational parameter of the Earth}, and $r_p$ is the distance between them at perigee.

When the distance at perigee equals the radius of Earth, the corresponding impact parameter $b$ is defined as $b_i$, representing the threshold value for a collision between the asteroid and Earth whenever $b\leq b_i$. Additionally, the perigee distance $r_p$ can be regarded as the deflection $\delta$. Consequently, if $\delta > 1$ Earth radii, the asteroid is predicted not to collide with Earth\footnote{This analysis disregards secondary effects such as atmospheric interaction.}.

\subsection{B-plane}
\label{sec:bplane}

{
The b-plane finds application in scenarios involving the approach of a small mass body toward a massive body. In the context of asteroid deflection, it is employed to characterize the asteroid's approach to Earth, acting as a link between heliocentric and planetocentric trajectories. This plane plays a crucial role in identifying potential collisions by calculating the impact parameter $b$ alongside its uncertainties and comparing it to $b_i$. Furthermore, the b-plane offers valuable insights by ``separating'' the effects of distance and phase in the analysis of the approach~\cite{letizia2016snappshot}.
}

\begin{figure}[!ht]
\centering
\includegraphics[width = .5\textwidth]{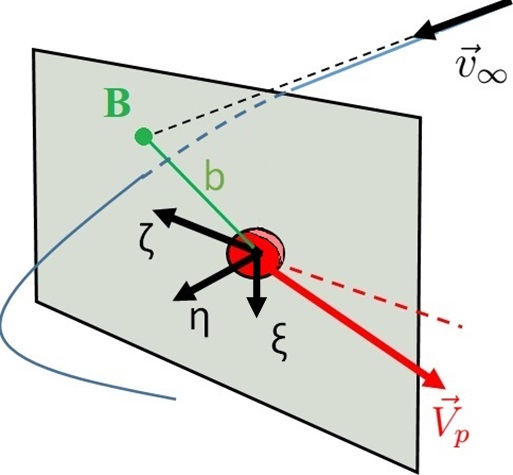}
\caption{Representation of the b-plane.}
\label{fig:bplane}
\end{figure}

Figure \ref{fig:bplane} illustrates this plane. It is defined to contain the center of mass of the planet with which the infinitesimal body makes the approach and is perpendicular to the relative velocity vector of the approach, represented in planetocentric coordinates as $\vec{v}_\infty$. Point B in Figure \ref{fig:bplane} represents the point where $\vec{v}_\infty$ would pierce the plane, serving as a reference for obtaining the impact parameter $b$. Moreover, a reference system is defined with its normal base formed by the following unit vectors:

\begin{subequations}
\begin{align}
\label{eqn:bplane1}
\hat{\eta} &= \frac{\vec{v}_\infty }{v_\infty}, \\
\label{eqn:bplane2}
\hat{\xi} &= \frac{\vec{V}_p \times \hat{\eta}}{||\vec{V}_p \times \hat{\eta}||}, \\
\label{eqn:bplane3}
\hat{\zeta} &= \hat{\xi} \times \hat{\eta},
\end{align}
\end{subequations}
where $\vec{V}p$ is the heliocentric velocity of the planet. The unit vector $\hat{\eta}$ is parallel to $v_\infty$, {while $\hat{\zeta}$ is parallel to the direction opposite to the projection of velocity $\vec{V}_p$ on the b-plane}, making it more sensitive to the phase between the planet and the approaching body. Finally, $\hat{\xi}$ completes the right-handed system and is more sensitive to the Minimum Orbit Intersection Distance (MOID).

As the impact parameter is contained in the b-plane, it can be easily obtained as:

\begin{equation}
\label{eqn:b_bplano}
b = \sqrt{\zeta^2+\xi^2}.
\end{equation}

Therefore, the deviation can be estimated as~\cite{sanchez2013impact}:

\begin{equation}
\label{eqn:desvio_bplano}
\delta=\frac{\mu_p}{v_\infty^2}\left(\sqrt{1+\left(\frac{b v_\infty^2}{\mu_p}\right)^2}-1\right),
\end{equation}
where $\mu_p$ represents the gravitational parameter of the planet.

\section{Circular Restricted n-Body Problem}
\label{sec:CRNBP}

{
To derive reasonably realistic trajectories for the selected group of asteroids, the simplified n-body model presented in Negri and Prado~\cite{negri2022nbody}, referred to as the circular restricted n-body problem (CRNBP), will be employed. This model will be complemented by the ephemerides correspondence elaborated in the same work, which is discussed in Section \ref{sec:CRNBP_efemeride}. The adoption of this simplified n-body problem is motivated by the need for a straightforward implementation with efficient processing times while ensuring alignment with the real orbital characteristics of the chosen asteroids.
}

\begin{figure}[!ht]
	\centering\includegraphics[width=.5\textwidth]{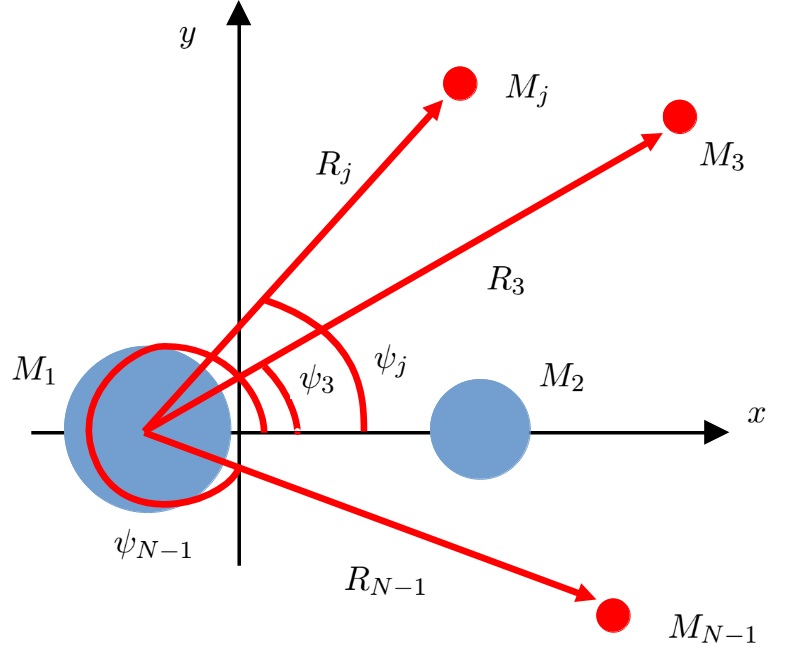}
	\caption{Representation of the CRNBP in the synodic frame.}
	\label{fig:CRNBP_Fig2}
\end{figure}	

The following system of equations represents the CRNBP for the primary bodies $M_1$ and $M_2$ in a synodic reference frame, in which $M_1$ and $M_2$ are fixed in the $x$-axis~\cite{negri2022nbody}:

\begin{subequations}
\label{eq:N2CRNBP}
\begin{align}
\begin{split}
\ddot{x} &= 2 \dot{y} + x - \frac{\mu_1}{r_1^3}(x+\mu_2) - \frac{\mu_2}{r_2^3} (x-\mu_1) - \sum_{j=3}^{N-1} \mu_j \left[ \frac{1}{r_j^3} (x+\mu_2-R_j \cos \psi_j ) + \right. \\ & \left.  \sum_{k=1,k\neq j}^{N-1} \frac{\mu_k}{(R_k^2+R_j^2-2 R_k R_j \cos(\psi_k-\psi_j) )^{3/2}} (R_j \cos \psi_j - R_k \cos \psi_k ) \right], 
\end{split} \\
\begin{split}
\ddot{y} &= - 2 \dot{x} + y - \frac{\mu_1}{r_1^3} y - \frac{\mu_2}{r_2^3} y - \sum_{j=3}^{N-1} \mu_j \left[ \frac{1}{r_j^3} (y-R_j \sin \psi_j ) + \right. \\ & \left.  \sum_{k=1,k\neq j}^{N-1} \frac{\mu_k}{(R_k^2+R_j^2-2 R_k R_j \cos(\psi_k-\psi_j) )^{3/2}} (R_j \sin \psi_j - R_k \sin \psi_k ) \right], 
\end{split} \\
\ddot{z} &= - \frac{\mu_1}{r_1^3} z - \frac{\mu_2}{r_2^3} z - \sum_{j=3}^{N-1}  \frac{\mu_j}{r_j^3} z. 
\end{align}
\end{subequations}
where $r_j$, for $j=1,2,...,N-1$, is the distance between the infinitesimal body and each planet. The mass parameter of each body is defined such that $\mu_j=M_j/(M_1+M_2)$. The $R_j$ represents the distance between each massive body and $M_1$, and $\psi_j$ is a phase angle defined from the x-axis of the synodic reference frame, as shown in Figure \ref{fig:CRNBP_Fig2}, for $j=1,2,...,N-1$, $k \neq j$ ($R_1=0$, $R_2=1$, and $\psi_2=0$). The chosen primaries are the Sun and the Earth, for $j=1$ representing the Sun, and $j=2$ the Earth. The arc $\psi_j$ is solved in time by the following analytical expression:

\begin{equation}
\psi_j = \psi_{0j}+ (n_j-n_{12}) t,
\end{equation}
where $\psi_{0j}$ represents the initial phase angle, $n_j$ is the mean motion of the $j$-th body in canonical units, and $n_{12}$ is the mean motion of $M_1$ and $M_2$ about their barycenter, which has a value of $n_{12}=1$ canonical units~\footnote{The mass is defined in canonical units such that $\mu_1 + \mu_2 = 1$, the distance is normalized so that the distance between the primaries is unity, and the time is normalized so that the rotational velocity of the synodic system is also unity ($n_{12}=1$)~\cite{szebehely1967theory,negri2022nbody}.}.

Note that this description of the problem reduces the 6n first-order differential equations of a restricted n-body problem to simply six differential equations and n-2 analytical expressions, making the analysis and simulation much simpler while considering the gravitational effects of all the planets. 

\subsection{Correspondence in Ephemerides}
\label{sec:CRNBP_efemeride}

Following the proposition made in Negri and Prado~\cite{negri2022nbody}, a correspondence between the ephemerides fixed system and the synodic frame can be obtained from:
\begin{subequations}
\begin{align}
\vec{\rho}_N = & T S \left( \vec{s}_{1N} - \mu_2 S^\text{T} T^\text{T} \hat{x} \right), \\
\dot{\vec{\rho}}_N = & \dot{T} S \left( \vec{s}_{1N} - \mu_2 S^\text{T} T^\text{T} \hat{x} \right) + T S \left( \dot{\vec{s}}_{1N} + \mu_2 S^\text{T} T^\text{ T} \dot{T} T^\text{T} \hat{x} \right),
\end{align}
\end{subequations}
where $\vec{\rho}_N = \begin{bmatrix} x & y & z \end{bmatrix}^\text{T}$ represents the position of the infinitesimal mass body in the synodic frame, $\vec{s}_{1N}$ is the position in the fixed inertial frame where the ephemerides are taken, here assumed as centered in the sun, and $\dot{\vec{s}}_{1N}$ the velocity.

The matrix $S$ is:
\begin{equation}
S = \frac{1}{\vert \vert \text{proj}_{\mathcal{O}_2} \vec{s}_{12}  \vert \vert} \begin{bmatrix}
 \text{proj}_{\mathcal{O}_2} \vec{s}_{12}^\text{T}  \\
(\hat{h}_2 \times \text{proj}_{\mathcal{O}_2} \vec{s}_{12})^\text{T} \\
\vert \vert \text{proj}_{\mathcal{O}_2} \vec{s}_{12}  \vert \vert \hat{h}_2^\text{T}
\end{bmatrix},
\end{equation}
in which the superscript $\text{T}$ represents the transpose, while $\hat{h}_2$ and $\text{proj}_{\mathcal{O}_2} \vec{s}_{12}$ obey:
\begin{subequations}
\label{eq:iygyhuh}
\begin{align}
    \hat{h}_2 &= \begin{bmatrix}\sin \Bar{i}_2 \sin \Bar{\Omega}_2 \\
    -\sin \Bar{i}_2 \cos \Bar{\Omega}_2 \\
    \cos \Bar{i}_2 \end{bmatrix},\\
    \hat{e}_2 &= \begin{bmatrix}\cos \Bar{\omega}_2 \cos \Bar{\Omega}_2 - \sin \Bar{\omega}_2 \sin \Bar{\Omega}_2 \cos \Bar{i}_2 \\
    \cos \Bar{\omega}_2 \sin \Bar{\Omega}_2 + \sin \Bar{\omega}_2 \cos \Bar{\Omega}_2 \cos \Bar{i}_2 \\
    \sin \Bar{\omega}_2\sin \Bar{i}_2 \end{bmatrix}, \\
    \hat{e}_{\perp 2} &= \hat{h}_2 \times  \hat{e}_2, \\
    \text{proj}_{\mathcal{O}_2} \vec{s}_{1j} &= (\vec{s}_{1j} \cdot \hat{e}_2) \hat{e}_2 + (\vec{s}_{1j} \cdot \hat{e}_{\perp 2}) \hat{e}_{\perp 2}
\end{align}
\end{subequations}
$\hat{h}_2$ is the mean unit vector of the angular momentum of $M_2$, and $\hat{e}_2$ is the mean unit vector of the Runge-Lenz-Laplace vector. The mean inclination, the argument of periapsis, and longitude of the ascending node of $M_2$ in the chosen fixed system are denoted as $\Bar{i}_2$, $\Bar{\omega}_2$, and $\Bar{\Omega}_2$, respectively.

The $T$ matrix transforms the rotating system at time $t=0$ to a time $t=t_N$ following:
\begin{equation}
\label{eq:CRNBP_T}
T = 
\begin{bmatrix}
\cos (n_{12} t_N) & \sin (n_{12} t_N) & 0 \\
-\sin (n_{12} t_N) & \cos (n_{12} t_N) & 0 \\
0 & 0 & 1
\end{bmatrix}.
\end{equation}
Note that $n_{12}=1$ when in canonical units and can be disregarded for subsequent steps.

The initial configuration of each massive body is:
\begin{equation}
\psi_{j0} = \arctan \left[ \frac{(\text{proj}_{\mathcal{O}_2} \vec{s}_{12} \times \text{proj}_{\mathcal{O}_2} \vec{s}_{1j})\cdot \hat{z}}{(\hat{h}_{2}\cdot \hat{z})({proj}_{\mathcal{O}_2} \vec{s}_{12} \cdot \text{proj}_{\mathcal{O}_2} \vec{s}_{1j})} \right].
\end{equation}

The mean motion of each of the bodies, $n_j$, is calculated from the average period of the orbit of each body in the fixed system of ephemerides.

Finally, considering that the ephemeris of the infinitesimal body is obtained at the epoch $JD$, defined in Julian date, $t_N$ can be obtained in canonical units as:

\begin{equation}
t_N = 86400(JD - JD_0) n_2,
\end{equation}
where $n_2$ is the mean motion of $M_2$, now in rad/s since its function here is to convert from seconds to canonical time units, and $JD_0$ is the selected Julian date for the correspondence of ephemeris for the other bodies. Note that this transformation is valid only if $t=0$ corresponds to the time when the ephemerides of the massive bodies are obtained. Therefore, to integrate the Equations \ref{eq:N2CRNBP}, with the initial condition of the ephemeris of the infinitesimal body, the initial time for the integration is $t_N$.

\section{Simplified Methods for Predicting Deflection}
\label{sec:simpli}

In this section, we present both methods that are likely to be applied in the preliminary design phase of the trajectory of a deflecting spacecraft. The approach presented in Section \ref{sec:izzo} is more suitable for low-thrust deflections, while the one presented in Section \ref{sec:vasile} will likely be the choice for impulsive approaches.

\subsection{Low-thrust semi-analytical approximation}
\label{sec:izzo}

{Izzo~\cite{izzo2007optimization}} proposes an approximation that considers the effect of a deflection to be so small that its impact is mainly on the mean motion, having a more significant effect on the phase, i.e., the instant when the asteroid intercepts Earth. Therefore, this approximation implies that the deflection has a greater impact on the variation of the $\zeta$ coordinate in the b-plane, thus~\cite{izzo2007optimization}:

\begin{equation}
\label{eq:anal_izzo}
\Delta \zeta = \frac{3a}{\mu_s}V_p\sin\vartheta\int_0^{t_p} (t_i - \tau) \vec{V} \cdot \vec{A} d\tau,
\end{equation}
where $\mu_s$ is the gravitational parameter of the Sun, $a$ denotes the semi-major axis of the asteroid's orbit, $\vartheta$ is the angle between the heliocentric velocity of Earth ($\vec{V}p$) and the asteroid's geocentric velocity as it enters Earth's sphere of influence ($\vec{v}_\infty$), $t_i$ is the interception time when the asteroid is intercepted and deflection begins, $\vec{A}$ is the acceleration resulting from the chosen deflection technique, and $t_p$ is the duration of $\vec{A}$'s action. Finally, $\vec{V}$ represents the heliocentric velocity of the asteroid.

Considering that $e$ is the eccentricity of the asteroid and $\theta$ the true anomaly, the magnitude of the asteroid's orbital velocity is:

\begin{equation}
\label{eq:V_mag}
V = \left[ \frac{\mu_S}{a(1-e^2)} (1+2e\cos \theta +e^2) \right]^{1/2}.
\end{equation}
Thus, Equation \ref{eq:anal_izzo} can be rewritten as:

\begin{equation}
\label{eq:anal_izzo2}
\Delta \zeta = 3 \sqrt{\frac{a}{\mu_S(1-e^2)}} V_p\sin\vartheta \int_0^{t_p} (t_i - \tau) (1+2e\cos \theta +e^2)^{1/2} A_t d\tau.
\end{equation}

Using Equation \ref{eqn:b_bplano}, we can find:

\begin{equation}
\label{eq:b_lowthrust}
b = \sqrt{ (\zeta_{MOID}+\Delta \zeta)^2 + \xi_{MOID}^2},
\end{equation}
where $\zeta_{MOID}$ and $\xi_{MOID}$ represent the b-plane coordinates of the MOID at the time of collision, i.e., the asteroid's coordinates on the b-plane if it was undeflected. With the magnitude of $b$ determined, we can utilize Equation \ref{eqn:desvio_bplano} to calculate the deflection with respect to the center of the Earth. {It is important to note that a new $v_\infty$ must be calculated. Since the approximation assumes that the main effects are on the phase of the encounter, a new $\theta_{MOID}$ can be computed by adding a time $\Delta t$ from the time of the undeflected asteroid impact with the Earth~\cite{izzo2007optimization}:
\begin{equation}
\Delta t = 3 \sqrt{\frac{a}{\mu_S(1-e^2)}}  \int_0^{t_p} (t_i - \tau) (1+2e\cos \theta +e^2)^{1/2} A_t d\tau.
\end{equation}
Subsequently, the Kepler equation is solved, taking into account this added $\Delta t$ to determine a new $\theta_{MOID}$, which can be easily incorporated with two-body relations to calculate a new $\vec{V}$ for the encounter of the asteroid and Earth, which is then used to obtain the new $v_\infty$.}

\subsection{Impulsive analytical approximation}
\label{sec:vasile}

The model proposed by Vasile and Colombo \cite{vasile2008optimal} is more general than the one by Izzo \cite{izzo2007optimization}, as it considers the perturbation in all orbital elements of the asteroid. However, it has the drawback that its simplicity applies only to impulsive techniques, requiring numerical integration in the case of low-thrust deflection.

The deflection in the MOID after a deflection is calculated by \cite{vasile2008optimal} as:

\begin{equation}
\vec{\Delta} = \begin{bmatrix}
\delta s_R & \delta s_T & \delta s_N
\end{bmatrix}^\text{T},
\end{equation}
where $\delta s_R$, $\delta s_T$, and $\delta s_N$ are the coordinates in radial-transverse-normal coordinates~\cite{vasile2008optimal}. They can be approximated to, as described in \cite{vasile2008optimal}:

\begin{subequations}
\begin{align}
\delta s_r \approx & \frac{R}{a}\delta a + \frac{ae\sin \theta_{MOID}}{\eta} \delta M - a \cos \theta_{MOID} \delta e, \\
\delta s_T \approx & \frac{R}{\eta^3} (1+e\cos \theta_{MOID})^2 \delta M + R \delta\omega + \frac{R \sin \theta_{MOID}}{\eta^2} (2+e\cos \theta_{MOID}) \delta e + R \cos i \delta \Omega,
\\
\delta s_N \approx & R (\sin \theta_{MOID}^*\delta i - \cos \theta_{MOID}^*\sin i \delta \Omega ),
\end{align}
\end{subequations}
where $e$, $i$, $\Omega$, and $\omega$ are the eccentricity, inclination, longitude of the ascending node, and argument of the perihelion of the asteroid, respectively. The true anomaly at the time of collision is $\theta_{MOID}$, and the latitude argument $\theta_{MOID}^*$ is calculated as $\omega + \theta_{MOID}$. Finally, the heliocentric distance in the nominal orbit is calculated as:

\begin{equation}
R = \frac{a(1-e^2)}{1+e\cos \theta_{MOID}}.
\end{equation}

It is considered that an impulse $\Delta \vec{V} = \begin{bmatrix} \Delta V_t & \Delta V_n & \Delta V_h \end{bmatrix}^\text{T}$ is applied to the asteroid, where the subscripts $t$, $h$, and $n$ represent the tangential, normal, and binormal coordinates~\cite{vasile2008optimal}, respectively.

Thus, the variation in each orbital element can be calculated using the Gauss Planetary Equations:

\begin{subequations}
\begin{align}
\label{eq:va_gauss}
\delta a \approx & \frac{2a^2V}{\mu_s}\Delta V_t, \\
\delta e \approx & \frac{1}{V} \left[ 2(e+\cos \theta_i)\Delta V_t - \frac{R}{a}\sin \theta_i \Delta V_n \right], \\
\delta i \approx & \frac{R \cos \theta^*_i}{h} \Delta V_h, \\
\delta \Omega \approx & \frac{R \sin \theta^*_i}{h \sin i} \Delta V_h, \\
\delta \omega \approx & \frac{1}{e V} \left[ 2 \sin \theta_i \Delta V_t + \left( 2e+\frac{R}{a}\cos \theta_i \right) \Delta V_n \right] - \frac{R \sin \theta^*_i \cos i}{h \sin i} \Delta V_h, \\
\label{eqn:deltaM}
\delta M \approx & - \frac{a\sqrt{1-e^2}}{eaV} \left[ 2\left( 1 + \frac{e^2R}{a(1-e^2)} \right) \sin \theta_i \Delta V_t + \frac{R}{a} \cos \theta_i \Delta V_n \right] + \delta n t_i,
\end{align}
\end{subequations}
where $\theta_i$ and $\theta^*_i$ represent the true anomaly and latitude argument at the moment of interception, respectively. The variation in the mean motion presented in Equation \ref{eqn:deltaM} is calculated as:

\begin{equation}
\delta n = \sqrt{\frac{\mu_s}{a^3}} - \sqrt{\frac{\mu_s}{(a+\delta a)^3}}.
\end{equation}

The deviation $\vec{\Delta}$ in the MOID can be expressed in the b-plane using Equations \ref{eqn:bplane1}, \ref{eqn:bplane2}, and \ref{eqn:bplane3}. Then, using Equation \ref{eqn:b_bplano}, we obtain:

\begin{equation}
\label{eq:b_impulsive}
b= \sqrt{(\zeta_{MOID}+\Delta_\zeta)^2+(\xi_{MOID}+\Delta_\xi)^2},
\end{equation}
{in which $\Delta_\zeta=\vec{\Delta} \cdot \hat{\zeta}$ and $\Delta_\xi=\vec{\Delta} \cdot \hat{\xi}$, for ``$\cdot$'' representing the dot product.}

Finally, with Equation \ref{eqn:desvio_bplano}, we can obtain the deflection $\delta$ achieved after the asteroid's deflection.

\section{Calculating the trajectory of impacting asteroids}
\label{sec:traj_impact}

In the following sections, {the methods presented previously will be applied to impacting asteroids}. However, it will be necessary to consider virtual asteroids since currently there are no asteroids with a high probability of colliding with Earth, according to the Center for Near Earth Object Studies (CNEOS) at NASA's Jet Propulsion Lab. For virtual asteroids, we mean asteroids for which the best estimate of their osculating elements, presented in an ephemerides table, is minimally adjusted to ensure the impact. It is important to note that this definition is more restrictive than the one used by centers like CNEOS, which define virtual asteroids as asteroids with a probability of impact higher than a certain threshold. Here, the definition is deterministic, meaning that the asteroid with a probability of an impact was selected, and its best-estimated orbit was minimally adjusted to achieve an impact in the utilized dynamical model (Section \ref{sec:CRNBP}). To justify the need for such a procedure, let's briefly discuss how the orbital determination of an asteroid and impact estimations are done.

The process of determining the orbit of a celestial body is similar to what is done with a spacecraft. However, the difference lies in the fact that for a newly discovered celestial body, which notably is not equipped with measuring instruments, the quality of observations tends to be highly inferior. The primary methods of observing an asteroid are through optical telescopes and radio telescopes. After obtaining a set of observations of the body, a linear regression using weighted least squares is applied to obtain the best estimate, considering the uncertainties of the observations \cite{farnocchia2015orbits}. Once the orbital determination is solved, the next step is to estimate the probability of the asteroid impacting Earth, which is done by propagating the estimate temporally with its associated uncertainties.

As discussed by Farnocchia et al. \cite{farnocchia2015orbits}, a heuristic rule is that for collision probabilities above 0.1$\%$, a linear model can be used. For probabilities below this threshold, the assumption of a Gaussian distribution becomes questionable. In this linear model, the estimate of the osculating elements is used in the reference trajectory for linearization and is propagated together with the covariance matrix from the orbital determination. With the covariance matrix, it is possible to determine a confidence ellipsoid around the reference trajectory at the time of closest approach to Earth. As expected, the validity of the linearity assumption diminishes when propagating over several years without new observations, relying on a limited number of observations, or dealing with observations featuring substantial uncertainties. In such instances, the adoption of nonlinear methods becomes essential, typically demanding considerable computational processing time.

One of these nonlinear methods is the Monte Carlo approach, in which the estimate from the orbital determination is varied in a normal distribution considering its covariance matrix. This process is repeated until a population of samples of size $N$ is formed. All samples are then integrated over the desired period. The collision probability is obtained by dividing the number of all integrations resulting in a collision by $N$. For more details on how to ensure reasonable confidence in the probability result, improve the robustness of the approach, and explore alternative nonlinear methods, refer to Farnocchia et al. \cite{farnocchia2015orbits}.

With this brief introduction to the calculation of the impact probability, it is evident that obtaining a realistically impacting trajectory from the data provided by CNEOS~\footnote{\url{cneos.jpl.nasa.gov/sentry/}} is extremely complex and would require deviations from the scope of this work. Here is why: very few asteroids would fit the heuristic rule for using a simple linear model. As of the consultation performed on the CNEOS website on December 28, 2021, only four asteroids have a probability of impact equal to or higher than 0.1$\%$. As a result, the use of nonlinear models would be necessary, which, in turn, would involve many simulations.

For example, applying a Monte Carlo approach to an asteroid with an impact probability $\mathcal{P}_{im}$ would require around $4/\mathcal{P}_{im}$ samples \cite{farnocchia2015orbits}. This implies numerous simulations, on the order of tens of thousands, millions, or even tens of millions. Furthermore, one would have to apply an n-body model with ephemerides and other perturbations (relativity, solar radiation pressure, etc.). Combining both factors turns the process into a Herculean effort in terms of implementation and computational processing time, deviating this work towards astrometry and compromising its objectives and scope. Moreover, in some cases, it is desirable to obtain a collision condition for an asteroid that is not predicted to impact, requiring a way of obtaining a virtual asteroid from the real asteroid.

Therefore, to obtain reasonably realistic trajectories for a group of selected asteroids, we will use the simplified n-body model presented in Section \ref{sec:CRNBP}, the CRNBP, with the corresponding ephemeris matching described in Section \ref{sec:CRNBP_efemeride}. The collision will be calculated using optimization methods. This approach is considered to satisfy the objectives of the studies to be presented in the following sections while maintaining some coherence with the selected real asteroid and keeping the implementation simple within a practical processing time.

The first step in the method used here to obtain the impacting trajectory is to define the impact epoch, $JD_{im}$, for the selected asteroid on the CNEOS website. Additionally, an arbitrarily chosen epoch of a month before the collision is selected to obtain the osculating elements of the asteroid from the ephemerides tables of the Horizons system of NASA's Jet Propulsion Lab~\footnote{\url{ssd.jpl.nasa.gov/horizons/}}, $JD=JD_{im}-1\text{ month}$. The ephemerides of the massive bodies, all the planets of the Solar System, used for the matching in the CRNBP (as described in Section \ref{sec:CRNBP_efemeride}), are selected for the impact epoch $JD_{im}$. The exception is for epochs of collision after January 9, 2100, due to the unavailability of Neptune's ephemerides after that date. In these cases, the correspondence in ephemerides of the massive bodies for January 1, 2100, is used. The fixed system of ephemerides is the ecliptic coordinate system, centered on the Sun, for the J2000 epoch.

In the CRNBP model, in addition to all the planets, we consider the Sun as $M_1$, as defined in Section \ref{sec:CRNBP}, and the Earth as $M_2$. The selected asteroids, along with some of their data obtained from the CNEOS website~\footnote{\url{cneos.jpl.nasa.gov/sentry/}}, such as the impact date (from which $JD_{im}$ can be obtained), and the estimated average diameter of the body, $D$, are presented in Table \ref{tab:select_ast_CNEOS}. {Asteroids without impact probability, $\mathcal{P}_{im}$, means there is no predicted impact with Earth, and the data considered for impact are for the closest approach to Earth, obtained from the Small-Body Database Lookup at NASA's Jet Propulsion Lab \footnote{\url{ssd.jpl.nasa.gov/tools}}.} Table \ref{tab:osc_horizons} shows the osculating orbital elements for each of these selected asteroids, as well as their respective epochs, $JD$, obtained from NASA's Jet Propulsion Lab Horizons system.

\begin{table}[!ht]
\renewcommand{\baselinestretch}{1.4}
\small
\centering
\caption{\label{tab:select_ast_CNEOS}Data of selected asteroids as impacting ones.}
\begin{center}
\begin{tabular}{lccc}
\hline
Asteroid&
Impact Date &
$\mathcal{P}_{im}$ [$\%$] &
$D$ [m] \\*
\hline
2005 ED224&
2023-03-11.35 &
0.0002 &
54 \\*
2008 EX5&
2072-10-09.02 &
0.0027 &
59 \\*
2008 UB7&
2063-11-01.40 &
0.00078 &
58 \\*
2009 JF1&
2022-05-06.34 &
0.026 &
13 \\*
2010 RF12&
2095-09-05.99 &
4.6 &
7 \\*
2015 JJ&
2111-11-07.26 &
0.0021 &
130 \\*
2020 VW&
2074-11-02.87 &
0.36 &
7 \\*
2021 EU&
2056-08-29.12 &
0.0034 &
28 \\*
3200 Phaethon &
2093-12-14.45 &
- &
6250 \\*
\hline
\end{tabular}
\end{center}
\renewcommand{\baselinestretch}{1.0}
\end{table}

After obtaining the necessary data, transformations are performed to convert the asteroid's ephemeris data to coordinates in the CRNBP's synodic system and canonical units. With this, an optimization problem is formulated, where the cost function to be minimized is given by:

\begin{equation}
\mathcal{J}(\vec{\Gamma}) = \delta = \left\Vert \vec{\rho}_N(t_{im},\vec{\rho}_{N0},t_N) - \vec{\rho}_2(t_{im}) \right\Vert.
\end{equation}
Here, maintaining coherence with Section \ref{sec:CRNBP}, $\vec{\rho}_N(t_{im},\vec{\rho}_{N0},t_N)$ indicates the position of the asteroid at the impact instant $t_{im}$, with the epoch $JD_{im}$ transformed into canonical units as defined in Section \ref{sec:CRNBP_efemeride}. It is also defined that $\vec{\rho}_{N0}=\vec{g}(\vec{\Gamma}): \mathbb{R}^6 \xrightarrow{} \mathbb{R}^3$, where $\vec{\rho}_{N0}$ represents the initial condition in the synodic system and canonical units for the initial time $t=t_N$ (obtained from $JD$, see Section \ref{sec:CRNBP_efemeride}), and $\vec{g}$ represents a function that transforms the vector of osculating orbital elements, $\vec{\Gamma}=\begin{bmatrix} a & e & i & \Omega & \omega & M\end{bmatrix}^\text{T}$, to the synodic system and canonical units. The vector $\vec{\rho}_2(t_{im})$ indicates the position of Earth at the instant $t_{im}$.

This first optimization problem aims to minimize the distance between the asteroid and Earth at the instant $JD_{im}$ using the CRNBP to guarantee the impact. To achieve this, a genetic algorithm is utilized, which performs a global search for solutions until the impact is ensured. Genetic algorithms are meta-heuristic optimization methods that, while not guaranteeing the global minimum, can explore a wide design space by utilizing biological concepts such as mutation, crossover, and selection. The \textit{ga()} function in MATLAB$^\circledR$ is used for this purpose. The optimization is typically performed multiple times until a solution is arbitrarily close to the osculating elements presented in Table \ref{tab:osc_horizons}, \ref{app:tables}, for the respective asteroid.

After the genetic algorithm stage is completed, the obtained solution or set of solutions is used as the initial guess for a second optimization problem. This optimization is a nonlinear programming problem with the following cost function to be minimized:

\begin{equation}
\label{eq:NLP_J}
\mathcal{J}(\vec{\Gamma})=\left\Vert W (\vec{\Gamma}^\star-\vec{\Gamma}) \right\Vert,
\end{equation}
where $\vec{\Gamma}^\star$ represents the osculating elements from Table \ref{tab:osc_horizons}, and $W \in \mathbb{R}^{6\times6}$ is a diagonal weighting matrix between the different orbital elements. Additionally, the following nonlinear constraint is imposed:

\begin{equation}
\label{eq:non_restr_delta}
\delta = \frac{1}{\rho_\earth} \left\Vert \vec{\rho}_N(t_{im},\vec{\rho}_{N0},t_N) - \vec{\rho}_2(t_{im}) \right\Vert \leq 1,
\end{equation}
where $\rho_{\earth}$ is the radius of Earth in canonical units, which means that the deviation $\delta$ in Equation \ref{eq:non_restr_delta} is represented in units of Earth radii.

Therefore, the condition $\delta \leq 1$ in the second optimization problem imposes the existence of a collision at the instant $t_{im}$. The new cost function in Equation \ref{eq:NLP_J} ensures a search for the virtual asteroid's orbital elements close to the real ones. For solving this optimization problem, the \textit{fmincon()} function in MATLAB$^\circledR$ is used, which employs an interior-point method (also known as barrier method). The diagonal weighting matrix $W$ is considered to have the following values: $W=\text{diag}\left(\begin{bmatrix}1000 & 1000 & 100 & 1 & 1 & 1\end{bmatrix} \right)$. The higher weight on the semi-major axis, eccentricity, and inclination is justified because these elements are the most useful for astronomers \cite{nesvorny2015identification} and are frequently used to classify and analyze asteroid families.

\begin{figure}[!ht]
\centering
\subfigure[3D View]{\includegraphics[width=.45\textwidth]{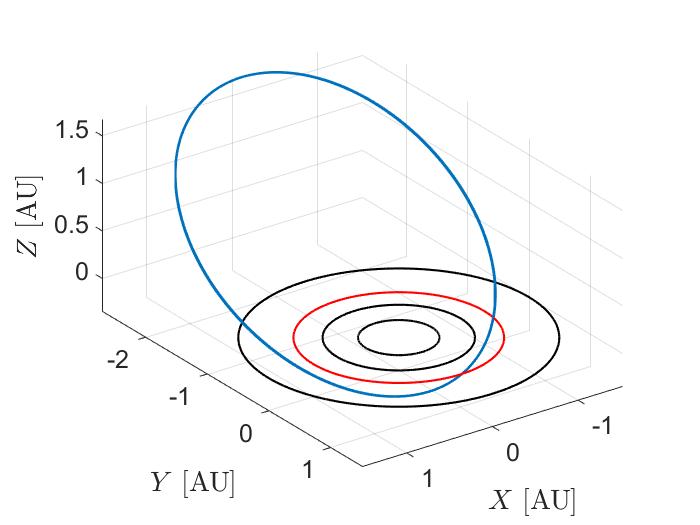}\label{fig_defl_traj_2005ED224}}
\subfigure[Ecliptic Projection]{\includegraphics[width=.45\textwidth]{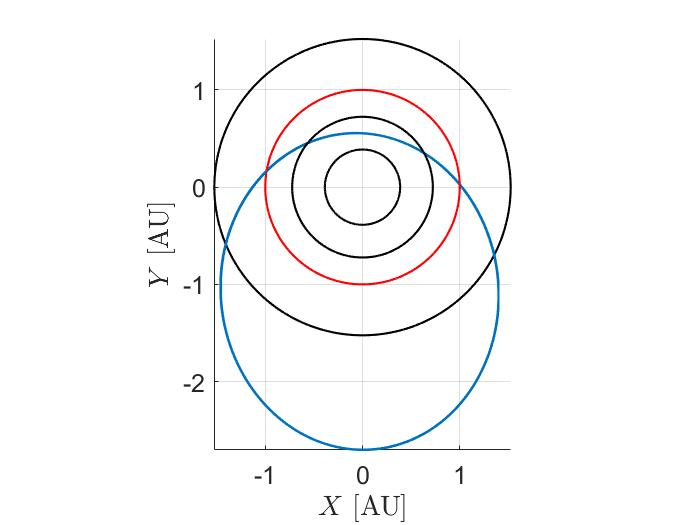}\label{fig_defl_proj_2005ED224}}
\caption{Trajectory of the virtual asteroid 2005 ED224.}
\label{fig_defl_2005ED224}
\end{figure}

The osculating elements of the virtual asteroids, which guarantee the impact with Earth while keeping minimal discrepancy from the real asteroid, at the end of the entire process, are presented in Table \ref{tab:osc_calculados}. The epochs remain the same as presented in Table \ref{tab:osc_horizons}. Figure \ref{fig_defl_2005ED224} shows the trajectory of the virtual asteroid 2005 ED224 up to 50 years before the collision with Earth in the fixed system. {The circular orbit of Earth in the CRNBP is represented in red, while Mercury, Venus, and Mars are shown in black.}

\section{Results and Discussion}
\label{sec:results}

As discussed in Section \ref{sec:intro}, the effect of other bodies on a deflection has been described in the literature, albeit only in limited aspects. Carusi et al. \cite{carusi2002} derived an analytical estimate based on Öpik's Theory to calculate the deflection of an asteroid due to an impulse. This estimate is very similar to Izzo's approach for impulsive deflection~\cite{izzo2007optimization}. By using this estimate and comparing it to an n-body problem, Carusi et al. show that the estimate describes well the deflection. However, according to their work \cite{carusi2002,carusi2005early,carusi2005mitigation,carusi2008orbital}, the divergence occurs when there are resonant encounters with the Earth between the time of interception and the predicted impact. In such cases, it is shown that the optimal impulse tends to diverge significantly from the analytical estimate, sometimes requiring a much larger or smaller impulse than expected.

As an example, consider their results (consult their work for context~\cite{carusi2002}) for the optimal impulse required to deflect the asteroid 1997 XF11 by one Earth radius, considering an n-body problem as the dynamical model. As shown by Carusi et al., their analytical model can predict the deflection with reasonable accuracy after 2028. However, there is a sharp increase in the required optimal impulse at the time of the 2028 interception, with a magnitude similar to the one needed to deflect the asteroid near the time of the predicted collision. Before 2028, the optimal impulse decreases by about two orders of magnitude. The events responsible for these divergences from the analytical model occur due to the asteroid's close approach to Earth in 2028, at a distance of about 0.005 AU, or about {0.8 SOIs (0.8 radius of the Earth's Sphere of Influence)\footnote{At times, SOI will be used as a unit of distance, indicating that the distance is measured in multiples of the radius of the sphere of influence of the respective planet. The definition of SOI used here is the one of Laplace \cite{vallado2001fundamentals}.}}.

Carusi et al.~\cite{carusi2002} show that the observed amplification in 2028 can be calculated analytically using the Extended Öpik Theory developed by Valsecchi and Carusi~\cite{valsecchi2003resonant}. This analytical formulation, although quite complex, relies on two-body approximations~\footnote{It uses the gravitational focusing presented in Section \ref{sec:FocGrav}, which depends on the approximation of a two-body problem.}, implying that it is only valid as an approximation for resonant passages within the sphere of influence of Earth. Moreover, as noted by Izzo et al.~\cite{izzo2006optimal}, the positions and shapes of gravitational keyholes are not well described by the analytical formulation, requiring numerical computations for an accurate description of the phenomenon.

We will demonstrate that the influence of a planet on the heliocentric trajectory of the asteroid, between the deflection moment and the predicted impact epoch, is much more extensive and complex than previously assumed in other works \cite{carusi2002,carusi2005early,carusi2005mitigation,carusi2008orbital}. General indications of this behavior were first observed in Negri et al.~\cite{negri2019third}, when analyzing the effects of Jupiter on a deflection using the BCR4BP. They were documented further by Negri~\cite{negri2022tese}, and shortly after by Chagas et al. ~\cite{chagas2022deflecting} (although this study considers only the Earth). Effects like those observed in Carusi et al.'s 1997 XF11 example do not only apply to Earth. Furthermore, and more importantly, they can occur at distances far beyond the perturbing body's sphere of influence, which can significantly affect the preliminary design phase of a deflecting spacecraft's trajectory, which mostly relies on two-body approximations. Understanding this phenomenon and knowing how to treat it increases the success of obtaining a set of reliable preliminary trajectories.

To achieve our goal, an n-body problem will be assumed, including all planets of the Solar System, and the approach from Section \ref{sec:traj_impact} will be used to calculate the deflection obtained, $\delta$, for a certain interception time $t_i$ before the collision. For the analytical model, the average of the osculating orbital elements of the undeflected asteroid over the period considered will be used. To calculate $\delta$ in the numerical model, it is sufficient to obtain the minimum distance of approach between the asteroid and Earth near the time of the collision. In the calculation of $\delta$ in the analytical model, it is necessary to consider $\zeta_{MOID}$ and $\xi_{MOID}$, as shown in Sections \ref{sec:izzo} and \ref{sec:vasile}. For this purpose, the $\zeta_{MOID}$ and $\xi_{MOID}$ obtained in the numerical model will be utilized. This choice is made to address the observation that the MOID tends to be significantly distant from Earth's orbit when using the average osculating orbital elements in the calculation of MOID through two-body approximations. In practice, this is done only to facilitate the comparison of models, as the purpose of a simplified model is not to achieve an extreme degree of precision but to provide a reliable estimate for preliminary and qualitative analyses.

\begin{figure}[!ht]
\centering
\subfigure[Impulsive deflection]{\includegraphics[width=.45\textwidth]{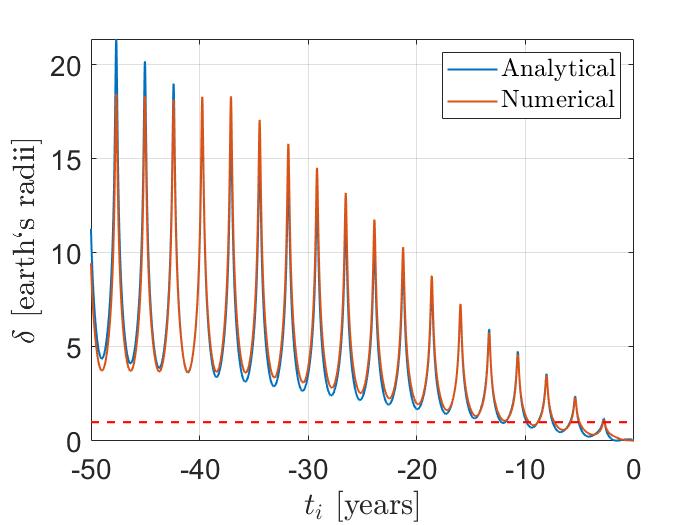}\label{fig_defl_impcomp_2005ED224}}
\subfigure[Low-thrust deflection]{\includegraphics[width=.45\textwidth]{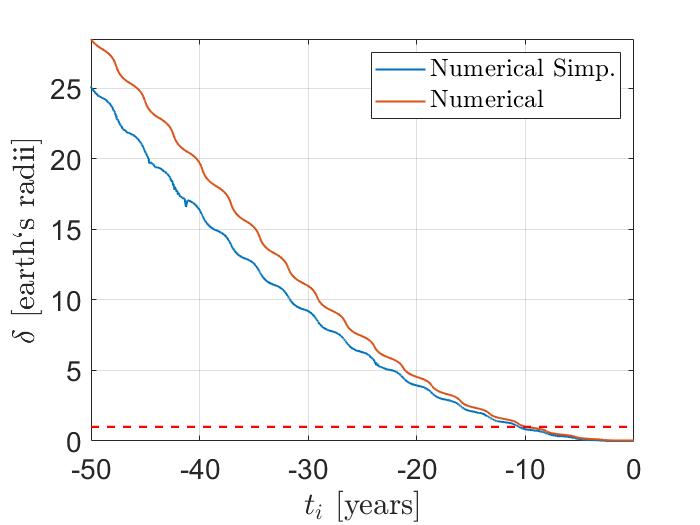}\label{fig_defl_lowcomp_2005ED224}} \
\subfigure[Errors]{\includegraphics[width=.45\textwidth]{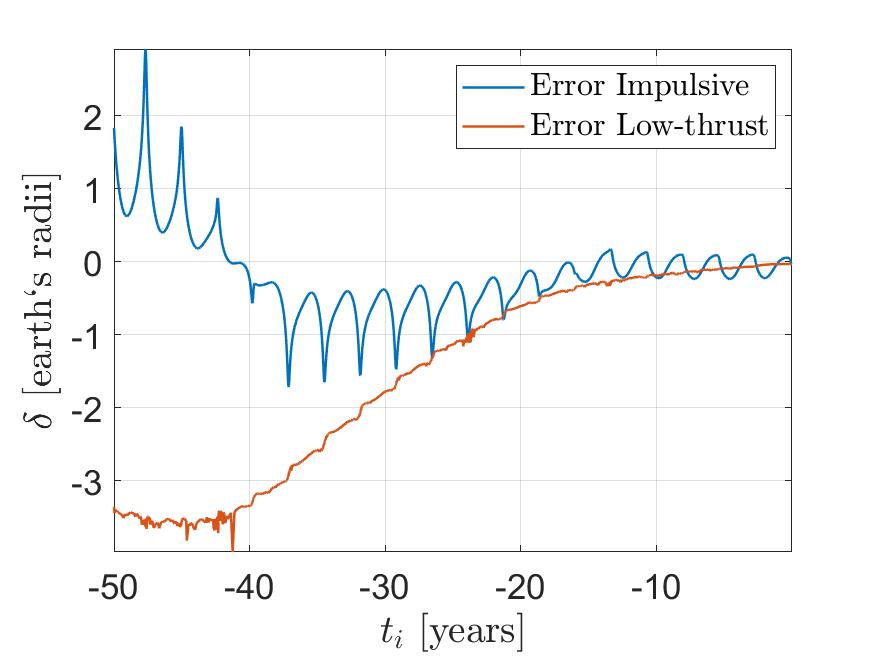}
\label{fig:errors_2005ED224}}
\subfigure[Distances]{\includegraphics[width=.45\textwidth]{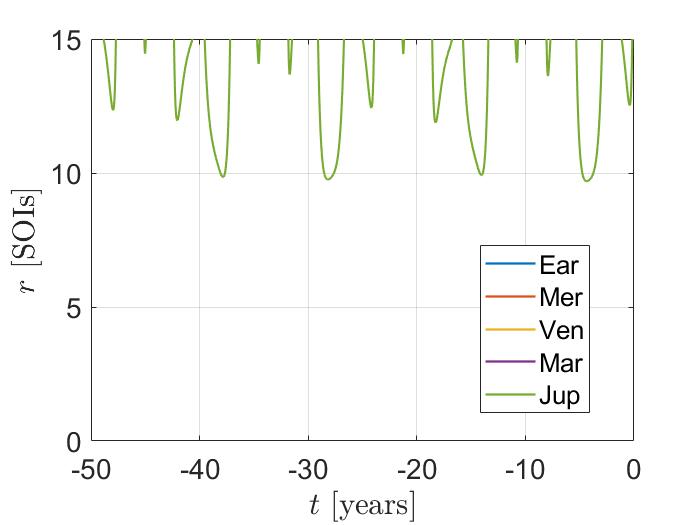}\label{fig_defl_apprcomp_2005ED224}}

\caption{Deflection obtained with $\Delta V =1$ cm/s or $A_t = 4\times 10^{-11}$ m/s$^2$ for asteroid 2005 ED224.}
\label{fig_defl_comp_2005ED224}
\end{figure}

\begin{figure}[!ht]
\centering
\subfigure[Impulsive deflection]{\includegraphics[width=.45\textwidth]{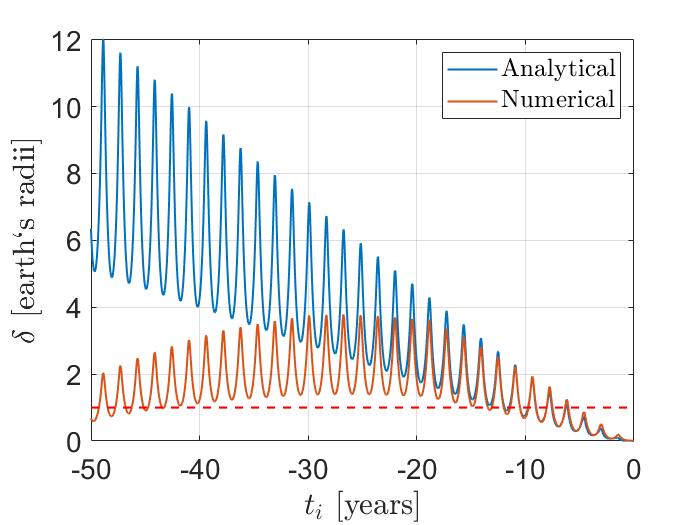}\label{fig_defl_impcomp_2008EX5}}
\subfigure[Low-thrust deflection]{\includegraphics[width=.45\textwidth]{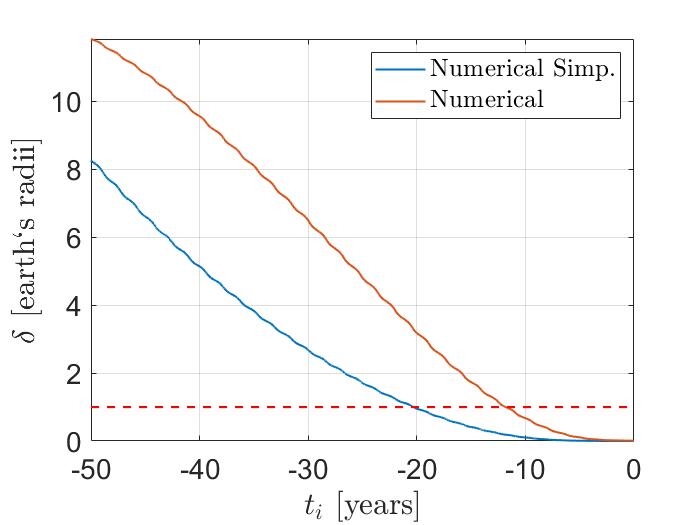}\label{fig_defl_lowcomp_2008EX5}} \
\subfigure[Errors]{\includegraphics[width=.45\textwidth]{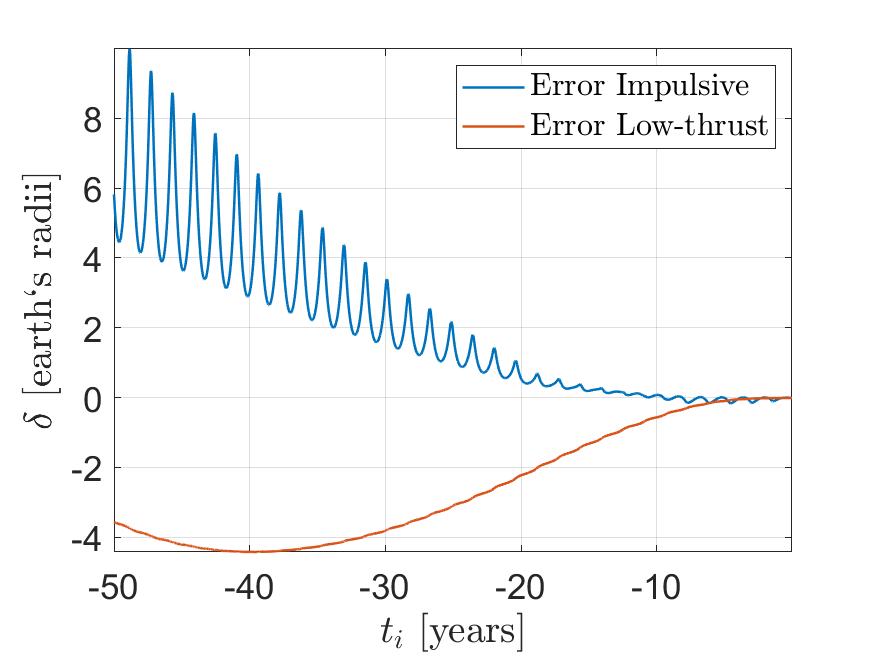}
\label{fig:errors_2008EX5}}
\subfigure[Distances]{\includegraphics[width=.45\textwidth]{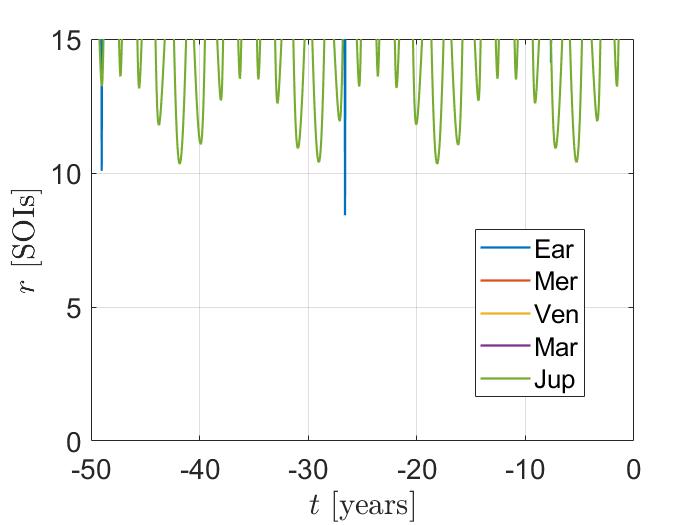}\label{fig_defl_apprcomp_2008EX5}}

\caption{Deflection obtained with $\Delta V =1$ cm/s or $A_t = 4\times 10^{-11}$ m/s$^2$ for asteroid 2008 EX5.}
\label{fig_defl_comp_2008EX5}
\end{figure}

\begin{figure}[!ht]
\centering
\subfigure[Impulsive deflection]{\includegraphics[width=.45\textwidth]{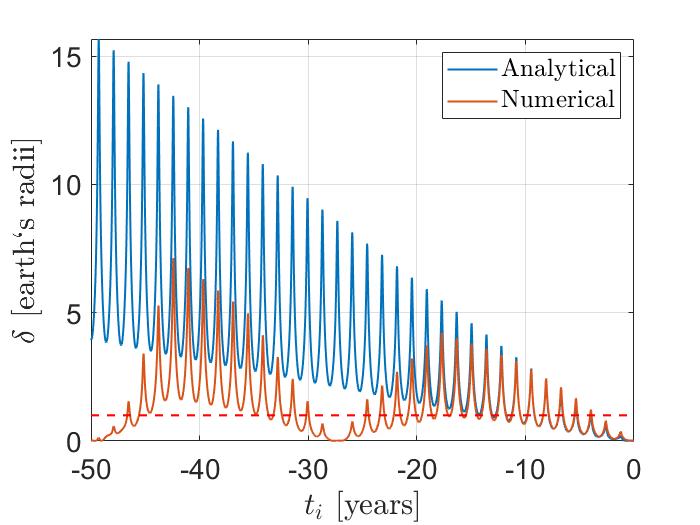}\label{fig_defl_impcomp_2008UB7}}
\subfigure[Low-thrust deflection]{\includegraphics[width=.45\textwidth]{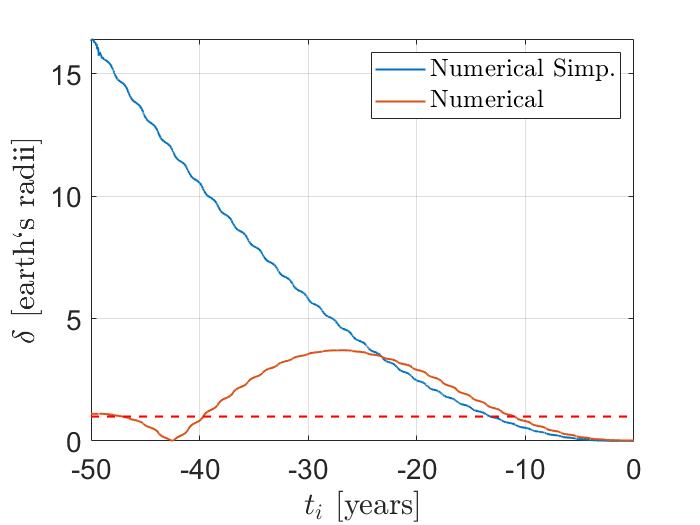}\label{fig_defl_lowcomp_2008UB7}} \
\subfigure[Errors]{\includegraphics[width=.45\textwidth]{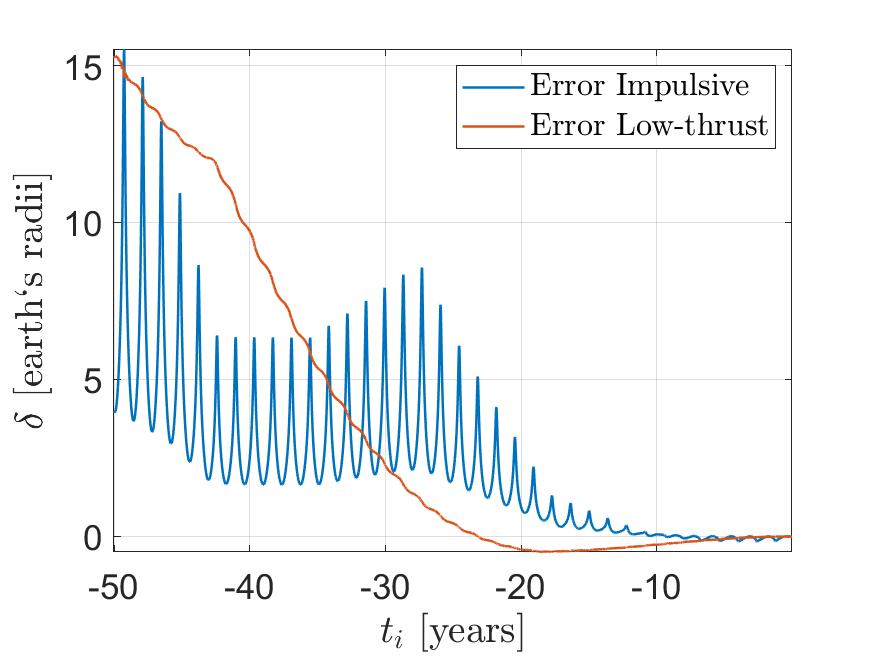}
\label{fig:errors_2008UB7}}
\subfigure[Distances]{\includegraphics[width=.45\textwidth]{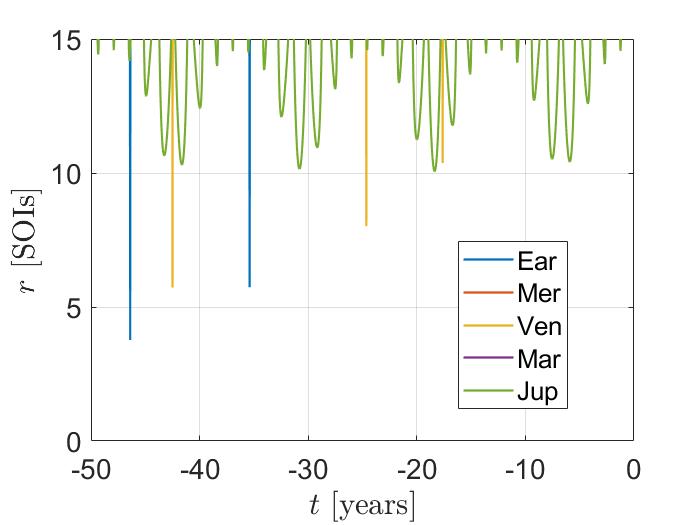}\label{fig_defl_apprcomp_2008UB7}}

\caption{Deflection obtained with $\Delta V =1$ cm/s or $A_t = 4\times 10^{-11}$ m/s$^2$ for asteroid 2008 UB7.}
\label{fig_defl_comp_2008UB7}
\end{figure}

\begin{figure}[!ht]
\centering
\subfigure[Impulsive deflection]{\includegraphics[width=.45\textwidth]{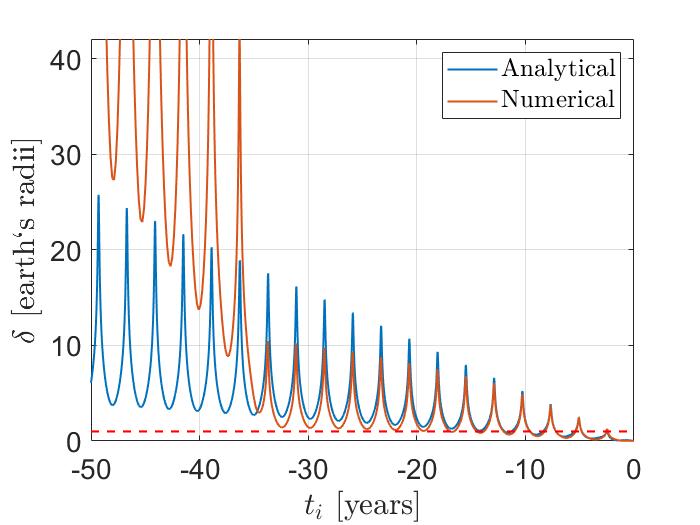}\label{fig_defl_impcomp_2009JF1}}
\subfigure[Low-thrust deflection]{\includegraphics[width=.45\textwidth]{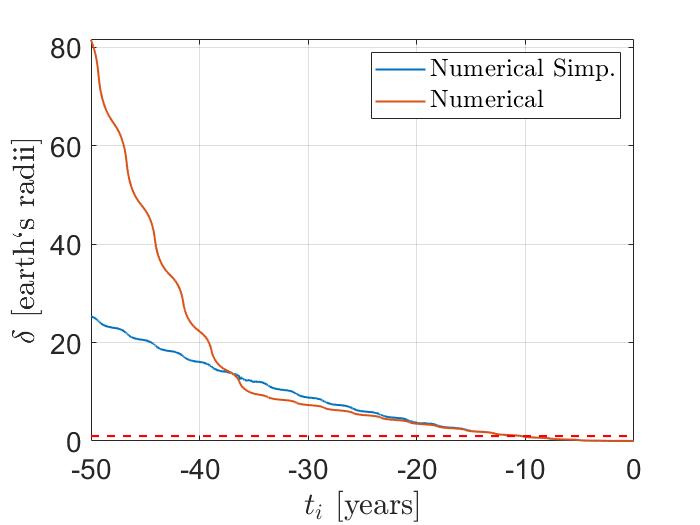}\label{fig_defl_lowcomp_2009JF1}} \
\subfigure[Errors]{\includegraphics[width=.45\textwidth]{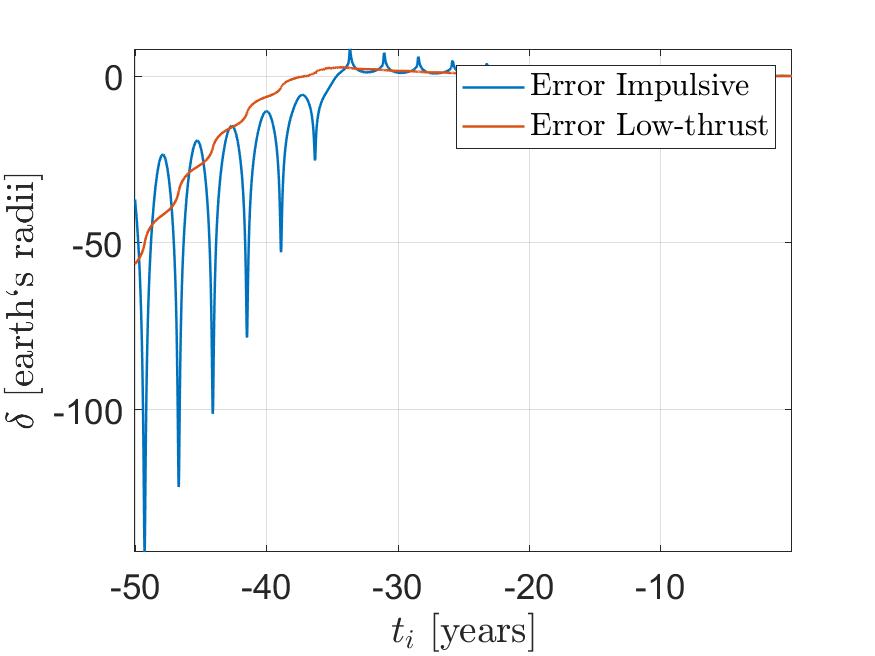}
\label{fig:errors_2009JF1}}
\subfigure[Distances]{\includegraphics[width=.45\textwidth]{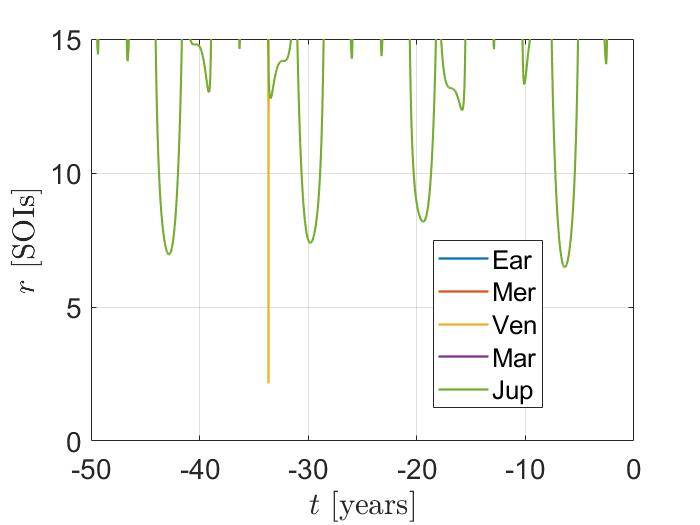}\label{fig_defl_apprcomp_2009JF1}}

\caption{Deflection obtained with $\Delta V =1$ cm/s or $A_t = 4\times 10^{-11}$ m/s$^2$ for asteroid 2009 JF1.}
\label{fig_defl_comp_2009JF1}
\end{figure}

\begin{figure}[!ht]
\centering
\subfigure[Impulsive deflection]{\includegraphics[width=.45\textwidth]{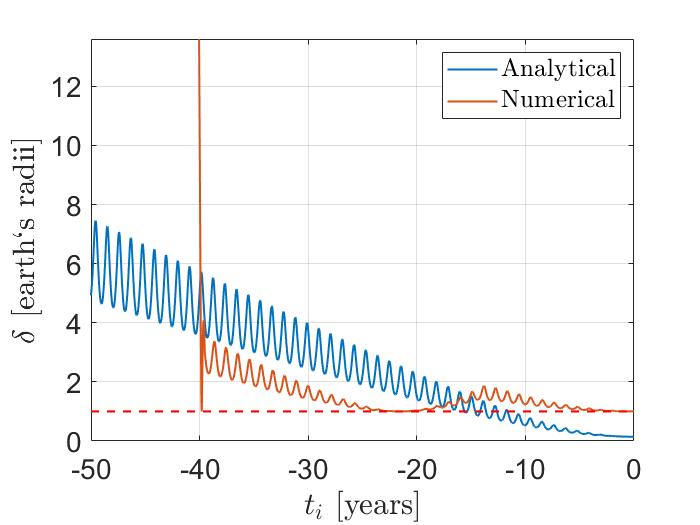}\label{fig_defl_impcomp_2010RF12}}
\subfigure[Low-thrust deflection]{\includegraphics[width=.45\textwidth]{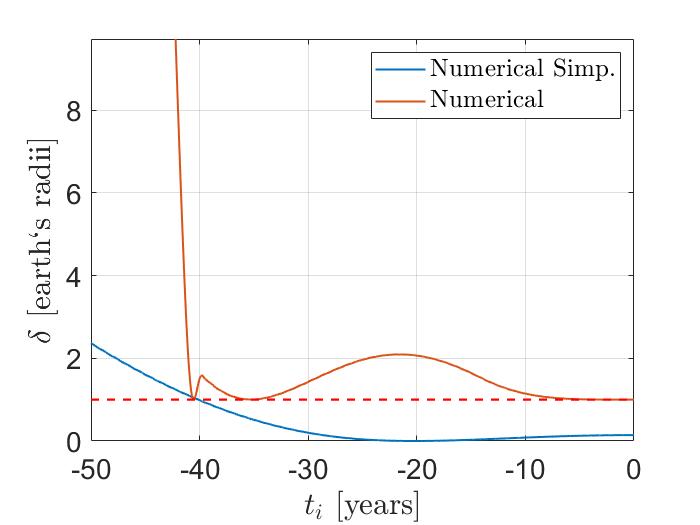}\label{fig_defl_lowcomp_2010RF12}} \
\subfigure[Errors]{\includegraphics[width=.45\textwidth]{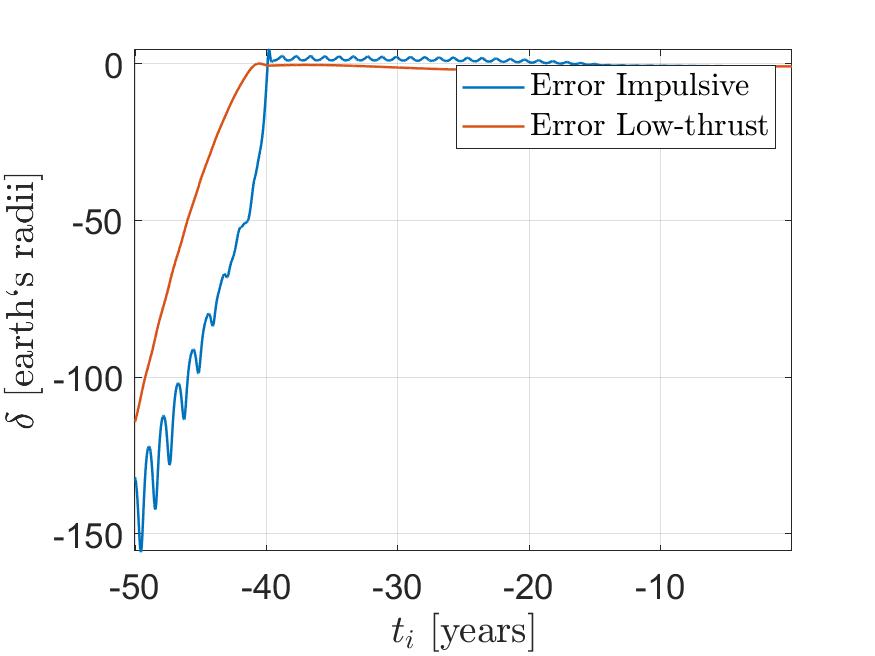}
\label{fig:errors_2010RF12}}
\subfigure[Distances]{\includegraphics[width=.45\textwidth]{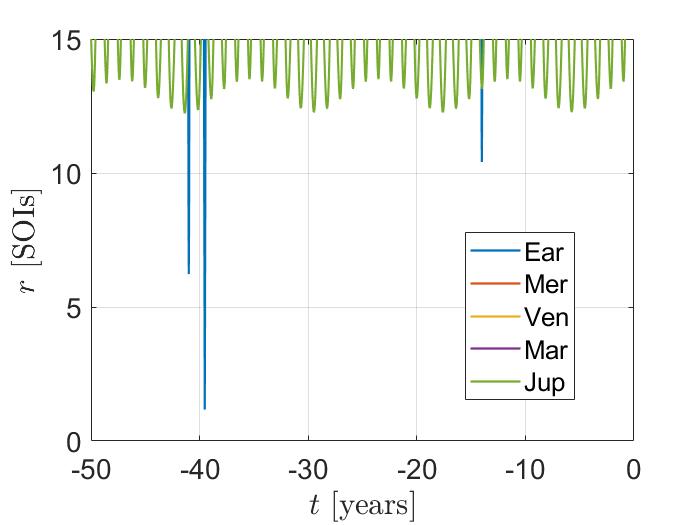}\label{fig_defl_apprcomp_2010RF12}}

\caption{Deflection obtained with $\Delta V =1$ cm/s or $A_t = 4\times 10^{-11}$ m/s$^2$ for asteroid 2010 RF12.}
\label{fig_defl_comp_2010RF12}
\end{figure}

\begin{figure}[!ht]
\centering
\subfigure[Impulsive deflection]{\includegraphics[width=.45\textwidth]{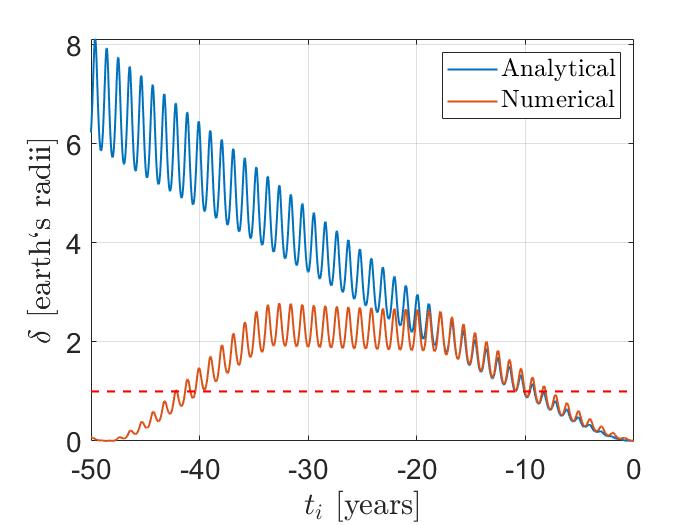}\label{fig_defl_impcomp_2015JJ}}
\subfigure[Low-thrust deflection]{\includegraphics[width=.45\textwidth]{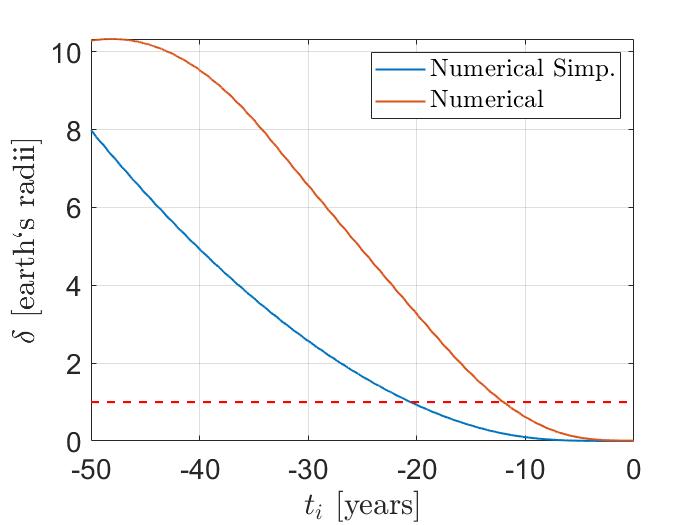}\label{fig_defl_lowcomp_2015JJ}} \
\subfigure[Errors]{\includegraphics[width=.45\textwidth]{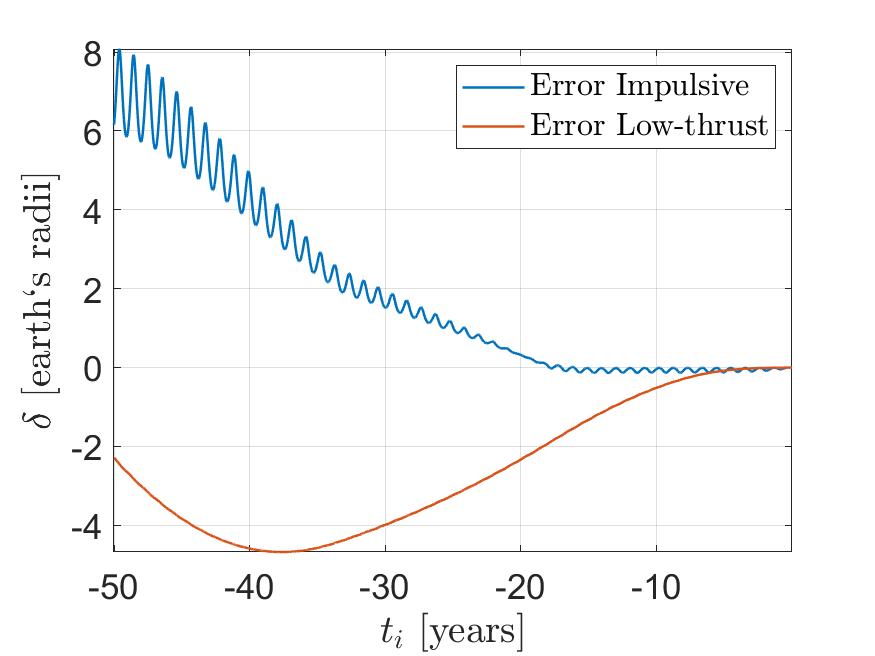}
\label{fig:errors_2015JJ}}
\subfigure[Distances]{\includegraphics[width=.45\textwidth]{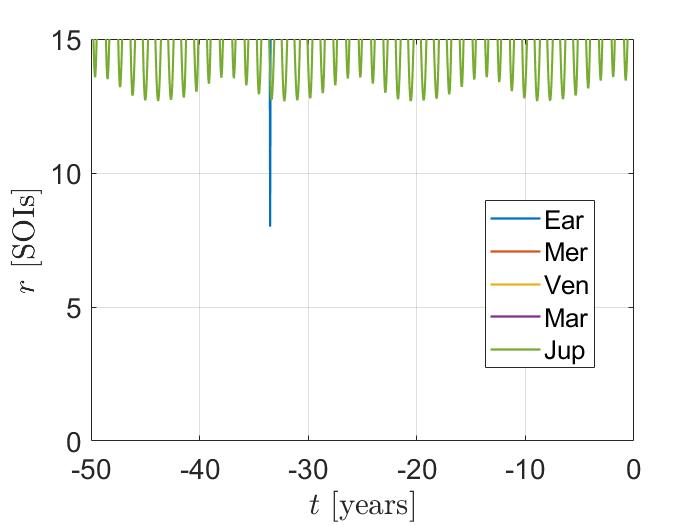}\label{fig_defl_apprcomp_2015JJ}}

\caption{Deflection obtained with $\Delta V =1$ cm/s or $A_t = 4\times 10^{-11}$ m/s$^2$ for asteroid 2015 JJ.}
\label{fig_defl_comp_2015JJ}
\end{figure}

\begin{figure}[!ht]
\centering
\subfigure[Impulsive deflection]{\includegraphics[width=.45\textwidth]{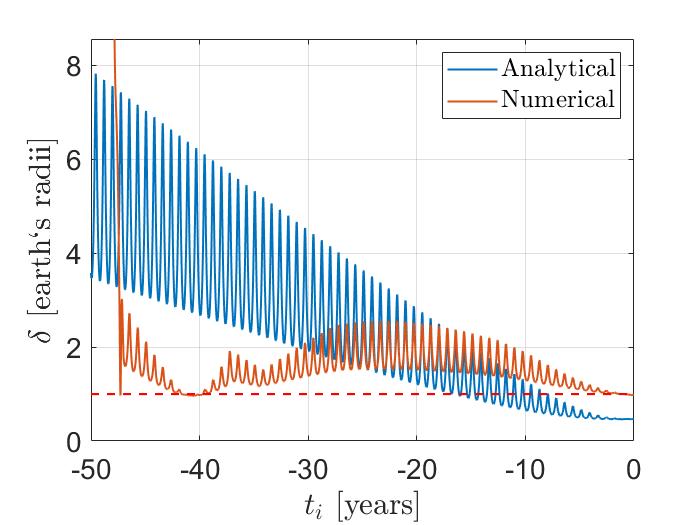}\label{fig_defl_impcomp_2020VW}}
\subfigure[Low-thrust deflection]{\includegraphics[width=.45\textwidth]{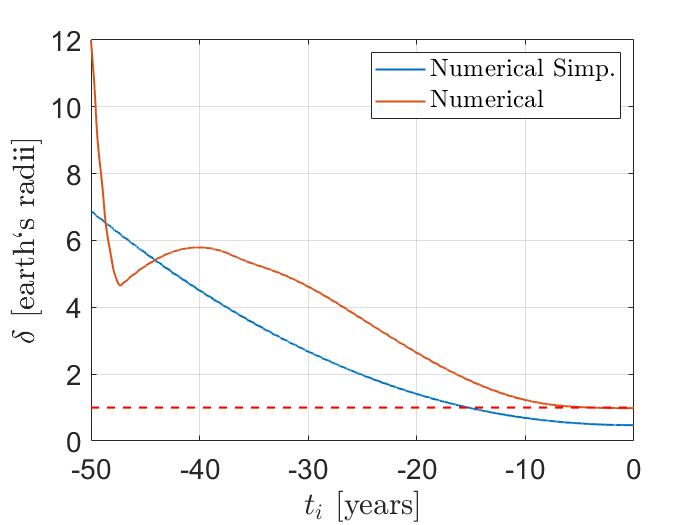}\label{fig_defl_lowcomp_2020VW}} \
\subfigure[Errors]{\includegraphics[width=.45\textwidth]{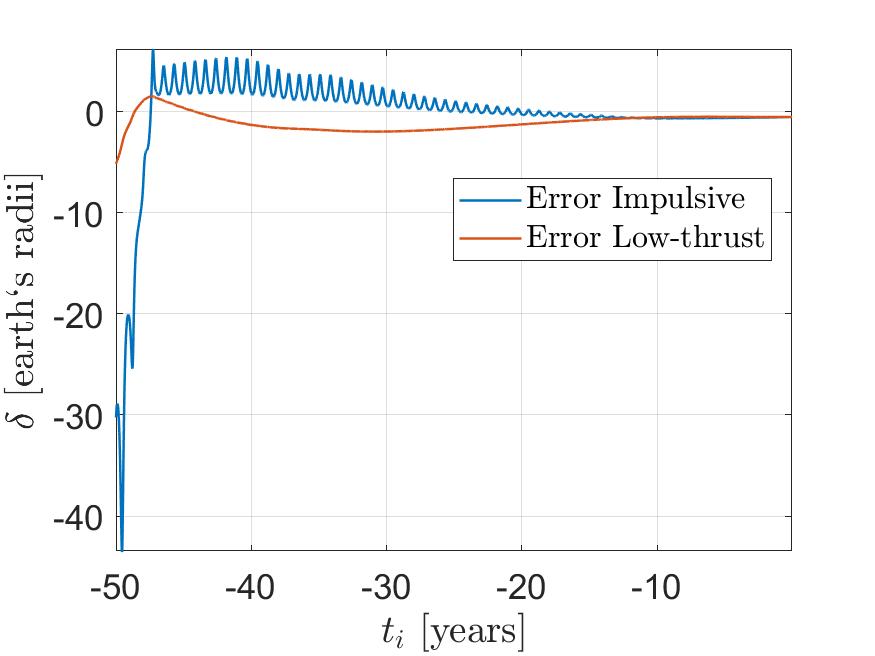}
\label{fig:errors_2020VW}}
\subfigure[Distances]{\includegraphics[width=.45\textwidth]{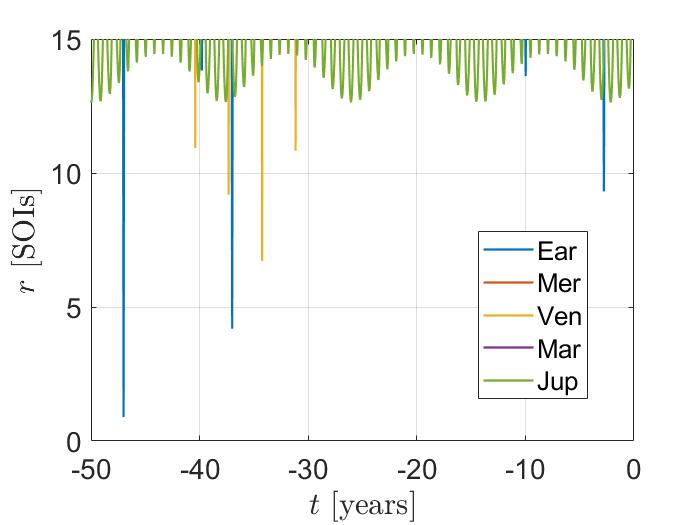}\label{fig_defl_apprcomp_2020VW}}

\caption{Deflection obtained with $\Delta V =1$ cm/s or $A_t = 4\times 10^{-11}$ m/s$^2$ for asteroid 2020 VW.}
\label{fig_defl_comp_2020VW}
\end{figure}

\begin{figure}[!ht]
\centering
\subfigure[Impulsive deflection]{\includegraphics[width=.45\textwidth]{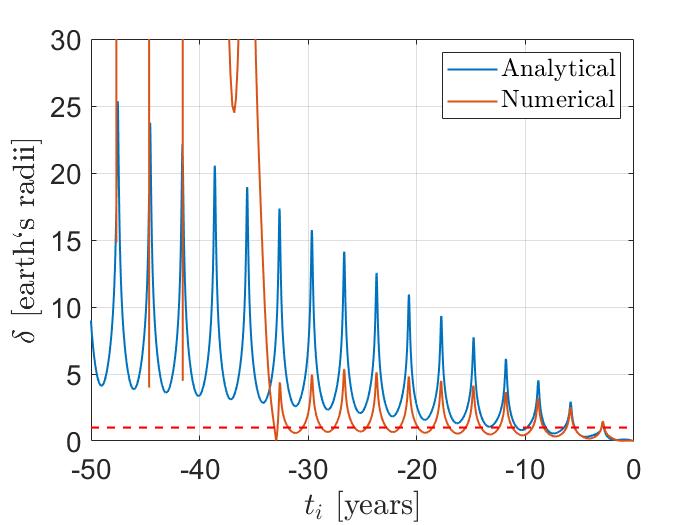}\label{fig_defl_impcomp_2021EU}}
\subfigure[Low-thrust deflection]{\includegraphics[width=.45\textwidth]{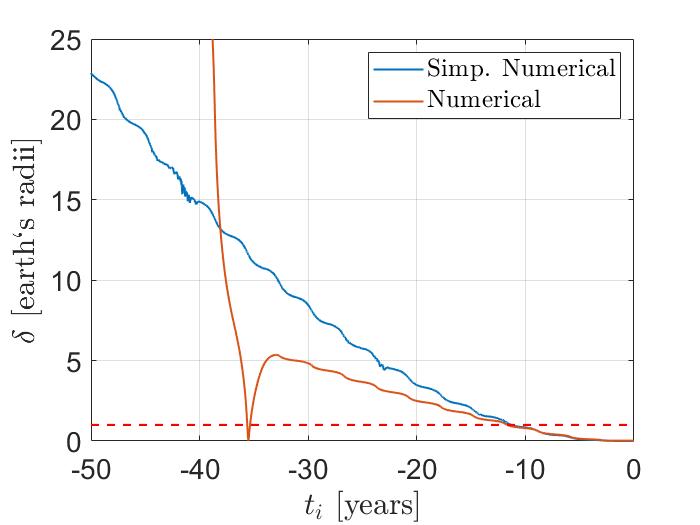}\label{fig_defl_lowcomp_2021EU}} \
\subfigure[Errors]{\includegraphics[width=.45\textwidth]{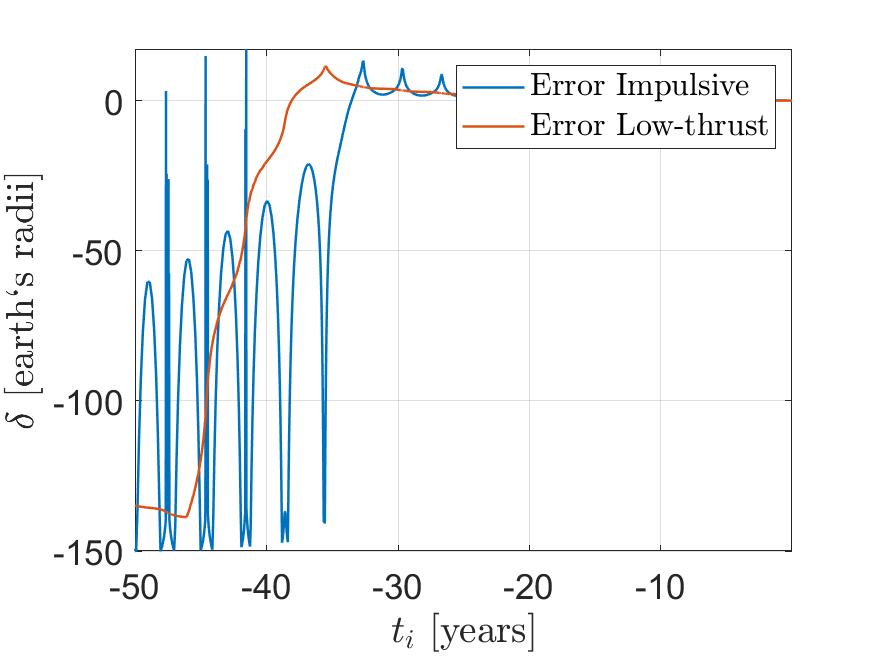}
\label{fig:errors_2021EU}}
\subfigure[Distances]{\includegraphics[width=.45\textwidth]{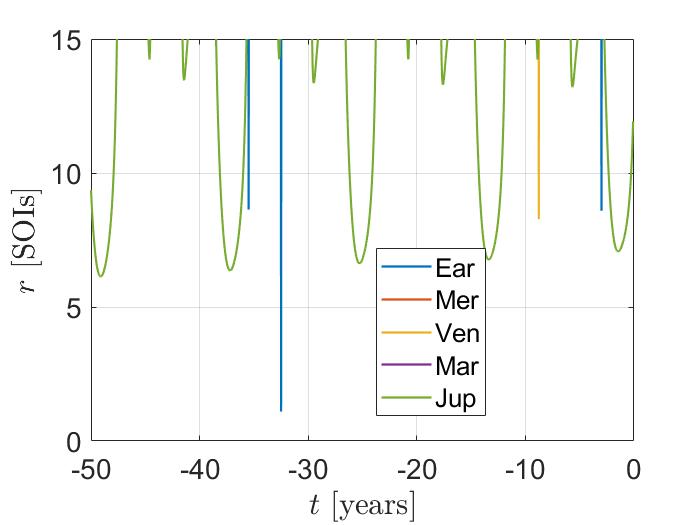}\label{fig_defl_apprcomp_2021EU}}

\caption{Deflection obtained with $\Delta V =1$ cm/s or $A_t = 4\times 10^{-11}$ m/s$^2$ for asteroid 2021 EU.}
\label{fig_defl_comp_2021EU}
\end{figure}

\begin{figure}[!ht]
\centering
\subfigure[Impulsive deflection]{\includegraphics[width=.45\textwidth]{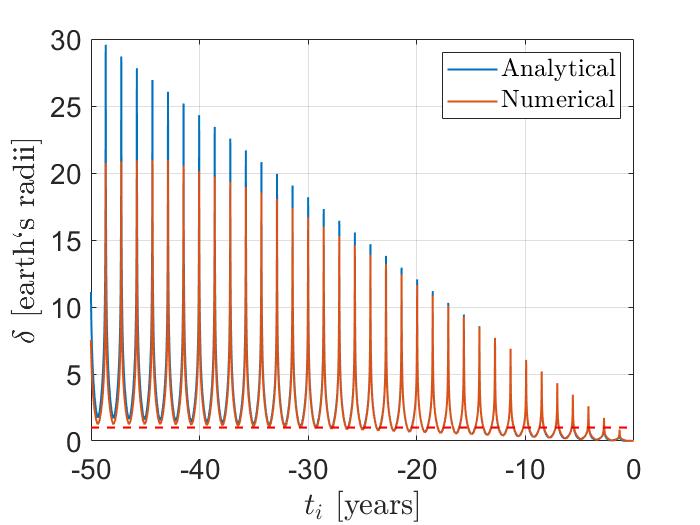}\label{fig_defl_impcomp_3200Phaethon}}
\subfigure[Low-thrust deflection]{\includegraphics[width=.45\textwidth]{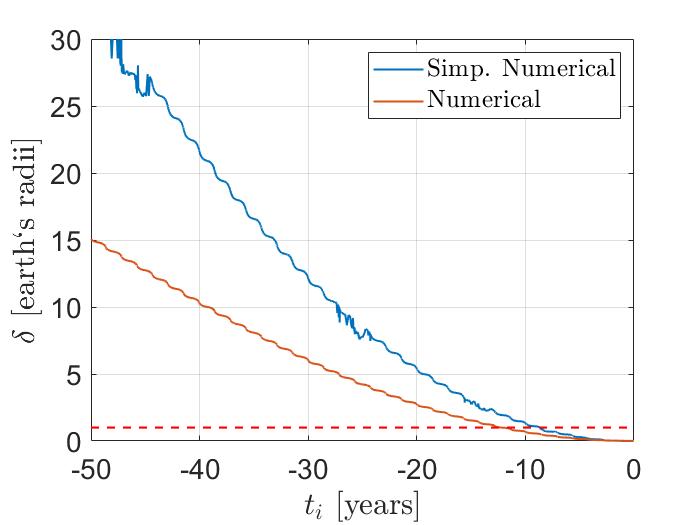}\label{fig_defl_lowcomp_3200Phaethon}} \
\subfigure[Errors]{\includegraphics[width=.45\textwidth]{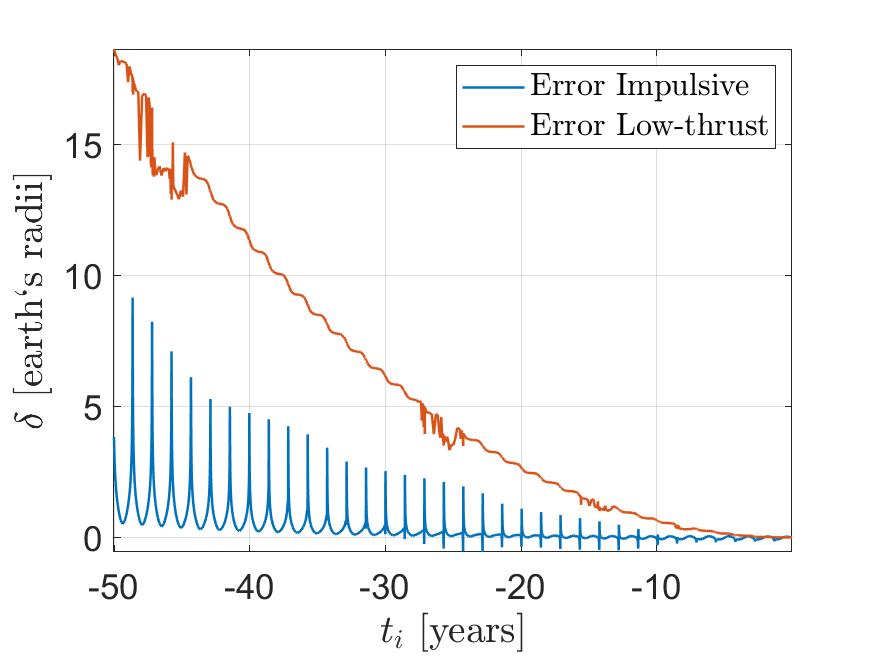}
\label{fig:errors_3200Phaeton}}
\subfigure[Distances]{\includegraphics[width=.45\textwidth]{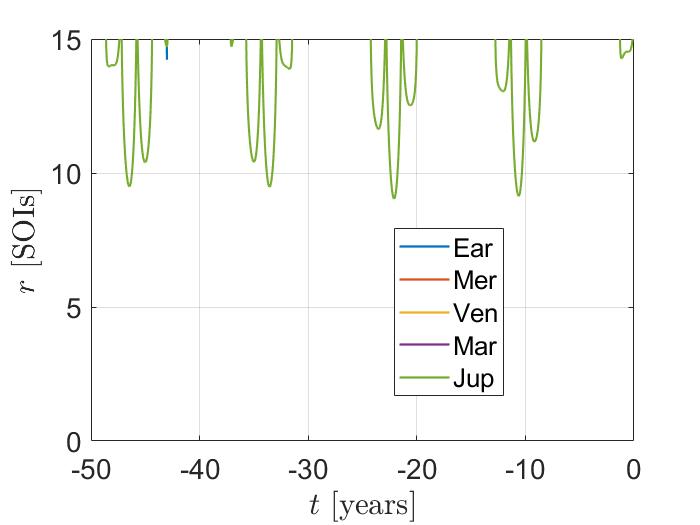}\label{fig_defl_apprcomp_3200Phaethon}}

\caption{Deflection obtained with $\Delta V =1$ cm/s or $A_t = 4\times 10^{-11}$ m/s$^2$ for asteroid 3200 Phaethon.}
\label{fig_defl_comp_3200Phaethon}
\end{figure}

As demonstrated in other works~\cite{park1999,ross2001,conway2001,park2003}, the optimal impulsive deflection tends to be dominant in the tangential component, and Izzo's approximation for low-thrust deflection (Section \ref{sec:izzo}) depends only on this same component. Therefore, the impulsive and low-thrust deflection models will be considered only in the tangential component, assuming an impulse of 1 cm/s and an acceleration of 40 pm/s$^2$~\footnote{The value of 40 picometers per second squared corresponds to a gravitational tractor of 10 tons at a distance of about 130 m from the asteroid's center of mass.}, respectively. Figures \ref{fig_defl_comp_2005ED224} to \ref{fig_defl_comp_3200Phaethon} show the results obtained for the different virtual asteroids considered here. 

The subfigures ``a'' represent the results obtained for the case of impulsive deflection (Section \ref{sec:vasile}), while itens ``b'' presents the case of low-thrust deflection (Section \ref{sec:izzo}). In these figures, the blue curves indicate the simplified methods of Section \ref{sec:simpli} while the orange curves indicate the results obtained with numerical integration {up to 50 years before impact with Earth}. The abscissas are represented by the time of interception ($t_i$), while the ordinates show the deflection obtained, in Earth radii, from a deflection beginning (low-thrust case) or at (impulsive case) $t_i$. {The dashed red line represents an Earth radius.  The errors between the simplified models and the numerical simulation are depicted in the subfigures ``c'', which is simply the numerical $\delta$ subtracted by the one obtained with the analytical models. The ``d'' subfigures indicate the distances to the planets of the Solar System along the trajectory of the undeflected asteroid, normalized by the sphere of influence (SOI) of each planet.}

Starting with impulsive deflection, as seen in Figures \ref{fig_defl_impcomp_2005ED224} to \ref{fig_defl_impcomp_3200Phaethon}, the obtained deflection shows significant oscillations depending on the interception time. These oscillations have a period equal to the asteroid's orbital period, with peaks representing deflections performed at the perihelion of the orbit, while valleys indicate deflection at the aphelion, already shown in previous works~\cite{park1999,ross2001,conway2001,park2003}. In the cases presented here, there is a wide range of results regarding the validity of the analytical model in Section \ref{sec:vasile} compared to numerical integration. In some cases, like Figures \ref{fig_defl_impcomp_2005ED224} and \ref{fig_defl_impcomp_3200Phaethon}, there is good agreement between the analytical prediction and the numerical result. In these cases, it is safe to say that an analytical model provides a reliable initial estimate, with qualitative and quantitative results quite close to those obtained from a more complex model. {For instance, it offers valuable initial estimates for more sophisticated optimization routines that consider the gravitational influence of other celestial bodies in subsequent stages of a deflecting spacecraft mission design.}

However, in some cases, such as Figures \ref{fig_defl_impcomp_2008EX5}, \ref{fig_defl_impcomp_2008UB7}, and \ref{fig_defl_impcomp_2015JJ}, the results obtained from integration diverge significantly from the analytical ones, both qualitatively and quantitatively. Nevertheless, there is at least agreement in the expected order of magnitude for the deflection. In other cases, some conditions lead to a huge difference in magnitude between the two methods, as shown in Figures \ref{fig_defl_impcomp_2009JF1}, \ref{fig_defl_impcomp_2010RF12}, \ref{fig_defl_impcomp_2020VW}, and \ref{fig_defl_impcomp_2021EU}. The same behavior applies to low-thrust deflection, as seen in Figures \ref{fig_defl_lowcomp_2005ED224} to \ref{fig_defl_lowcomp_3200Phaethon}.

The important discovery of this research is the explanation of these divergences, which invalidate the analytical models in certain applications. These discrepancies are related to a single reason, namely, the asteroid's approximation to perturbing bodies during or after the deflection. More broadly than observed in previous works, these divergences do not depend solely on resonant encounters with Earth. Any planet can invalidate the analytical model if the asteroid approaches it at a sufficient distance. The aggravating factor is that the sufficient proximity that would cause such divergence can be considerably far from the perturbing body, on the order of a dozen times the radius of the body's sphere of influence, {making the application of simple approximations challenging (if at all possible)}. {We will refer to these encounters as ``shallow encounters''}. 

This behavior is illustrated in Figures \ref{fig_defl_apprcomp_2005ED224} to \ref{fig_defl_apprcomp_3200Phaethon}, which show the distance from the non-deflected asteroid to each of the planets, from Mercury to Jupiter, normalized by the radius of the sphere of influence (SOI) of the respective body. {Table \ref{tab:dist_enc} complements these figures by showing the planet encountered by the asteroid at the time $t$ and the distance of the closest approach. Table \ref{tab:dist_enc} only shows encounters within 20 SOIs, and it omits the encounters with Jupiter because it is too numerous.}

{
Before delving into this aspect, it is noteworthy that certain cases already exhibit divergence very close to impact, as illustrated in Figures \ref{fig_defl_impcomp_2010RF12} and \ref{fig_defl_impcomp_2020VW}. This is due to the incongruity of using $\zeta_{MOID}$ and $\xi_{MOID}$ from the ``real data'' (in this study, the numerical result of the virtual asteroid's trajectory) with $\Delta \zeta_{MOID}$ and $\Delta \xi_{MOID}$ obtained from the analytical models in Eqs. \ref{eq:b_lowthrust} and \ref{eq:b_impulsive}. In both cases, there are many approaches to a planet relatively close to the impact. See Table \ref{tab:dist_enc} for both asteroids. These approaches are enough to change considerably the asteroid's orbit close to the impact. However, as the approach is close to the impact, this ``new'' asteroid's orbit (after the encounter) has less significance in the calculation of the mean orbital elements, which are used to compute $\Delta \zeta_{MOID}$ and $\Delta \xi_{MOID}$ with the simplified models, than its ``former orbit''. However, the ``mean orbit'' right before the impact is what defines the geometry of the impact. In other words, the used mean orbital elements are far enough from the ``mean orbit'' right before the impact, slightly miscalculating $\Delta \zeta_{MOID}$ and $\Delta \xi_{MOID}$, while  $\zeta_{MOID}$ and $\xi_{MOID}$ depicts exactly the impact geometry for this new orbit because they are taken from the numerical results.

As already argued, using an analytical estimation of $\zeta_{MOID}$ and $\xi_{MOID}$ is not an option since it will end up with estimations very far from the Earth, making the problem worse. Taking the mean over the osculating elements in the ``new'' asteroid's orbit (after the approach) is not an option either. That is because these other mean orbital elements will represent a better trajectory of the asteroid after the approach, but not before it. As we intend to study the effects of shallow planetary encounters, it did not seem prudent to fix this small incongruity in aggravating the divergence of the analytical to the numerical results previous to the encounter. Moreover, as seen in Figures \ref{fig_defl_impcomp_2010RF12} and \ref{fig_defl_impcomp_2020VW}, the predicted deflection behavior remains the same, only with a small shift (no more than a single radius of the Earth) in the curve. For this work, there is no need to overcomplicate the method in trying to remove that shift just to address these two particular cases.
}

{Returning to the discussion on shallow encounters, a straightforward comparison of Figures \ref{fig_defl_apprcomp_2005ED224} and \ref{fig_defl_apprcomp_3200Phaethon} with their respective subfigures ``a'' and ``b'' clearly reveals significant divergences in the simplified models when deflection takes place earlier or during the shallow encounter between the asteroid and a planet.} Take, for example, the asteroid 2008 UB7, Figure \ref{fig_defl_comp_2008UB7}, the agreement between the simplified models and the numerical one is reasonable until around $t_i \approx -18$ years, when there is an approach to Venus at just over 10 SOIs~\footnote{Specifically, the encounter is at a distance of 10.4 SOIs at $t=-17.6$ years. See Table \ref{tab:dist_enc}.}. Approaches between 2008 UB7 and Earth and Venus for $t_i$ before -25 years contribute to make the scenario even more discrepant and complex. These interactions lead to conditions of collision in impulsive deflection around $-30 \lessapprox t_i \lessapprox -25$ and $t_i \lessapprox -45$ years, whereas the analytical model predicts a deflection by a few terrestrial radii.

In the case of low-thrust deflection, impact with Earth occurs for $t_i \lesssim -42$ years if the spacecraft maintains thrust tangential to the asteroid's velocity throughout the period. This has significant operational implications for low-thrust deflection methods. For instance, in the case of a gravity tractor or an ion beam deflection, the deflecting spacecraft may need to switch between various hovering flight stations to fulfill its mission, instead of continuously maintaining the station in the tangential direction. For deflection techniques {with no active control}, such as those involving modification of the asteroid's thermal-optical properties, this highlights the need for a high degree of confidence in all properties of the body and all involved dynamics and emphasizes the necessity to account for these perturbations in highly complex numerical simulations.

As observed, there are instances where the order of magnitude of the deflection undergoes considerable changes. These cases typically occur at distances closer to the perturbing body. However, it would be incorrect to assume that this change necessarily happens within the planet's Sphere of Influence (SOI). The case of asteroid 2009 JF1 illustrates this point well. An approach to Venus at more than 2 SOIs is sufficient to significantly amplify the deflection effect, as depicted in Figure \ref{fig_defl_comp_2008UB7}. Among the cases presented here, the only entry into the SOI occurs for asteroid 2020 VW when it comes at 0.9 SOIs of Earth, as shown in Figure \ref{fig_defl_apprcomp_2020VW} and Table \ref{tab:dist_enc}.

For asteroids 2008 EX5, Figure \ref{fig_defl_comp_2008EX5}, and 2015 JJ, Figure \ref{fig_defl_comp_2015JJ}, a discrepancy arises between the models even though there does not seem to be an approach between the asteroid and any perturbing body at the beginning of the divergences. In reality, these are encounters not shown in the figure or that are difficult to visualize. There is an approach of 18.5 SOIs between 2015 JJ and Earth at $t_i \approx -18$ years, {as presented in Table \ref{tab:dist_enc}}. For 2008 EX5, there is an encounter with Earth at $t_i \approx -7.6$ years at a distance of approximately 14.1 SOIs. The choice not to represent approaches greater than 15 SOIs in the figures is to maintain their clarity. There is a reasonable number of approaches between 15 and 20 SOIs that do not cause considerable divergence between the simplified and numerical models. {Nevertheless, they are represented in Table \ref{tab:dist_enc}.} In the exceptions where there is a divergence for such approximations, which among the presented asteroids is only the case for 2015 JJ, they are peculiarities of the approach geometry.

To understand the influence of the approach geometry, one must resort to the circular restricted three-body problem (CR3BP), and studies on the dynamics of a gravity-assist, such as those developed by Qi and Xu~\cite{qi2015mechanical} and Negri and Prado~\cite{Negri:2018:AnLiAp}. As demonstrated by Qi and Xu~\cite{qi2015mechanical}, it is possible to write the rate of change of energy near the second primary in a CR3BP as follows:

\begin{equation}
\label{eq:dEdt}
\frac{dE}{dt} = y \mu (1-\mu) \left( \frac{1}{r_1^3} - \frac{1}{r_2^3} \right),
\end{equation}
where $y$ is the coordinate on the y-axis of the synodic system of the CR3BP, $r_1$ is the distance to the most massive body, and $r_2$ is the distance to the second primary of the CR3BP.

\begin{figure}[!ht]
\centering
\subfigure[Projection]{\includegraphics[width=.45\textwidth]{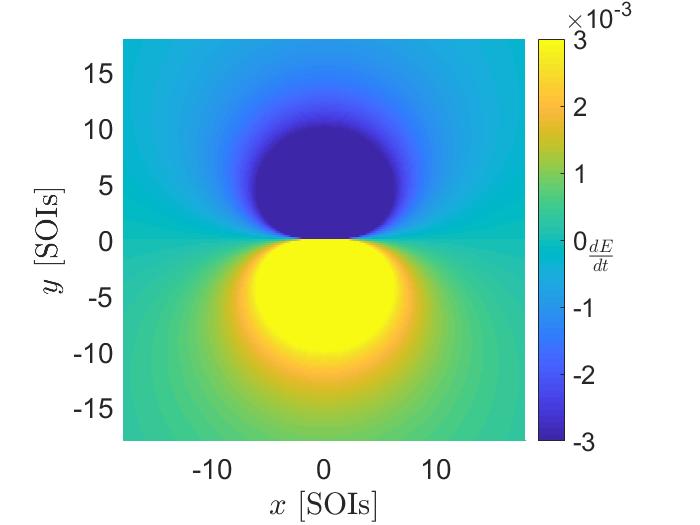}\label{fig_dEdt_proj}}
\subfigure[Cuts]{\includegraphics[width=.45\textwidth]{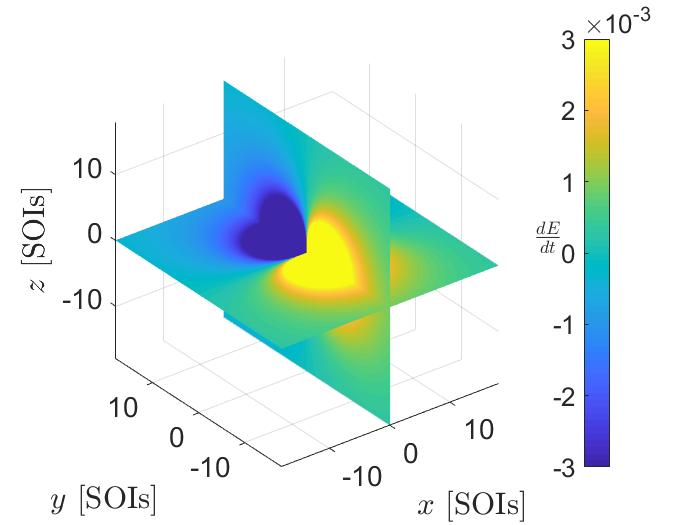}\label{fig_dEdt_3d}} \\
\subfigure[Illustrative Exaggerated Example]{\includegraphics[width=.35\textwidth]{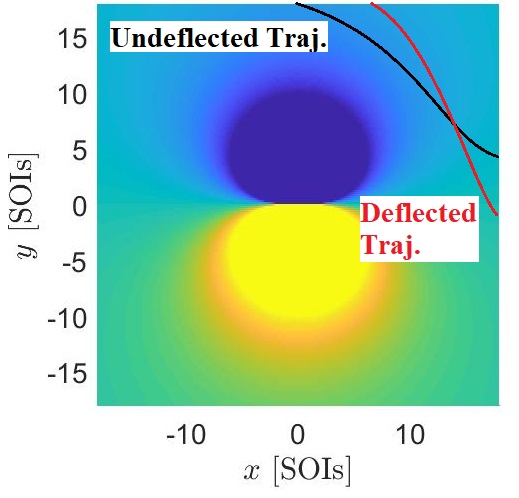}\label{fig_dEdt_illus}}
\caption{Energy variation near the planet Venus.}
\label{fig_dEdt}
\end{figure}

Figure \ref{fig_dEdt} shows the curves of $dE/dt$ for the cuts $x=0$ and $z=0$, applying Equation \ref{eq:dEdt} to a CR3BP between the Sun and Venus. In this figure, the synodic system is transposed from the barycenter to the center of Venus, so that $\vec{r}_2 = \begin{bmatrix} x  & y & z \end{bmatrix}$. As discussed by Qi and Xu~\cite{qi2015mechanical} and Negri and Prado~\cite{Negri:2018:AnLiAp}, the energy variation is positive for passages in the {trailing edge of} Venus and negative for passages in the {leading edge of} Venus. Here, ``trailing edge'' is defined as the region of negative $y$; that is, where Venus is coming from in its orbit. On the other hand, the term ``leading edge'' refers to regions of positive $y$, where Venus is moving to in its orbit. By keeping the distance $r_2$ constant, it can be noted that the greatest energy variations occur along the $y$ axis when $x=0$ and $z=0$ in this synodic system centered on Venus.

{
The deflection applies a small perturbation to the heliocentric trajectory of the asteroid. This small perturbation slightly displaces the trajectory in the shallow encounter, compared to the undeflected trajectory, as exemplified in Fig. \ref{fig_dEdt_illus} by an illustrative exaggerated displacement. The small displacement, which is hardly noticeable in a plot when using real data, causes the deflected asteroid to pass in slightly different values in the contours of $dE/dt$. When this change in the encounter geometry results in a net energy change of the same order of magnitude as or greater than the energy added by the deflection~\footnote{Although energy is just one of the integrals of motion in a two-body problem, its behavior can be extended to the other integrals. Thus, discussing energy is sufficient to exemplify the effect of the asteroid's approach to a perturbing body after a deflection.}, it invalidates the analytical model. In other words, even though the approach with the planet is quite distant, the small displacement of the trajectory tends to ``collect a perturbation'' to the same order of magnitude as or more than the deflection. 

{The relative velocity is crucial in this context, as a lower relative velocity implies that the deflected asteroid will spend more time accumulating a different level of perturbations during the shallow encounter in the slightly displaced trajectory. This principle is well-established and has been observed in the design of multiple gravity-assists in outer planetary systems~\cite{Negri:2018:AnLiAp}}. Those are the reasons why it is difficult to make a good simplified approximation to describe such a phenomenon, since the whole encounter trajectory should be considered, comparing the net effects of the deflected to the undeflected orbit in the encounter.
}

\begin{figure}[!ht]
\centering
\includegraphics[width = .6\textwidth]{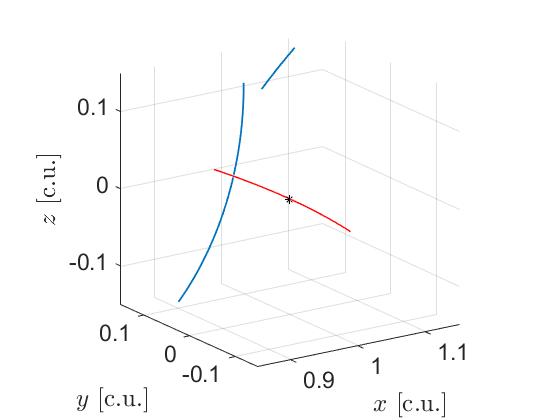}
\caption{shallow encounter at $~18.5$ SOIs that greatly impacted the prediction of deflection utilizing the analytical models for the asteroid 2015 JJ.}
\label{fig:Fig_2015JJ_geometry_approach}
\end{figure}

{
Now, considering the case of the asteroid 2015 JJ. The divergence between the models for approach distances greater than 15 SOIs only occurs for specific geometries of the encounter, generally happening in regions more accentuated in the $dE/dt$ magnitude shown in Fig. \ref{fig_dEdt_3d}, with greater perturbation of the trajectory. Figure \ref{fig:Fig_2015JJ_geometry_approach} shows this specific shallow encounter between 2015 JJ and Earth in the synodic frame~\footnote{The run to obtain this figure was considered up to $t=-20$ years. That is why it does not present the closer encounter at $t=-33.5$ years and $~8$ SOIs of distance.}. The black asterisk represents the Earth (which is fixed in this frame), the red line represents the Earth's orbit, and the blue line is the asteroid's trajectory. The uppermost section of the trajectory is a shallow encounter happening at $~-16.5$ years and $~29$ SOIs of distance. The remaining trajectory's section is the shallow encounter at $-18$ years and $~18.5$ SOIs of distance (see Table \ref{tab:dist_enc}), which is responsible for perturbing the deflected asteroid's orbit enough to create the divergence between the analytical and numerical computations in Fig. \ref{fig_defl_comp_2015JJ}. 

Note that the asteroid in this encounter passes through the area with the larger magnitudes of Fig. \ref{fig_dEdt_3d}, right in front of the Earth. Therefore, to accurately determine if the approach to the perturbing body is sufficient to invalidate any of the simplified models, one would need to find the trajectory's geometry relative to the body, making it relatively difficult to check the analytical models' validity when such encounters happen.

To document these encounters and their geometry of approach, we introduce an angle $\alpha$, defined as follows:

\begin{equation}
    \cos \alpha = \frac{y}{r_2},
\end{equation}
Here, $y$ and $r_2$ correspond to the vector $\vec{r}_2 = \begin{bmatrix} x  & y & z \end{bmatrix}$, which is utilized to generate Fig. \ref{fig_dEdt} (\underline{not} in Eq. \ref{eq:dEdt}). We calculate $\alpha$ at the point of closest approach. Practically, this angle signifies how close or distant the closest approach is from the $y$-axis in the frame centered on the planet, where a more significant perturbation occurs, as illustrated in Fig. \ref{fig_dEdt}. 

Table \ref{tab:dist_enc} presents the values of $\alpha$ in each shallow encounter. In cases where $\alpha>90^\circ$, a new angle $\alpha^\star=\alpha-180^\circ$ is provided in the table to better represent approaches occurring on the trailing edge, denoted with a negative sign. It is essential to note that the angle $\alpha$ offers an intuitively condensed view of the approach. However, larger values of $\alpha$ do not necessarily imply an absence of perturbation, as the entire encounter trajectory must be considered, not just the closest approach. The same holds for small values; they indicate that the spacecraft has passed through high $dE/dt$ regions. Yet, once again, the net perturbation depends on the entire encounter trajectory and the cumulative effects induced by the deflection. The angle $\alpha$ serves as an indicative measure.

In the case of asteroid 2015 JJ, the shallow encounter at $-18$ years and approximately $18.5$ SOIs of distance resulted in an angle $\alpha$ of $8.9^\circ$, aligning with the observations in Fig. \ref{fig:Fig_2015JJ_geometry_approach}. The angle $\alpha$ sheds light on the behavior depicted in Fig. \ref{fig:errors_2005ED224}. While, as previously discussed, we can assert that the errors in the case of 2005 ED224 remain qualitatively and quantitatively reasonable, likely providing good initial estimates, a noticeable change in error behavior is apparent for the impulsive comparison in the range of $-40 \leq t_i \leq -35$ years. Examining Table \ref{tab:dist_enc}, we find an encounter with Earth at -37 years, at a distance of 16.3 SOIs, with an associated angle $\alpha$ of $-10.7^\circ$. This indicates that the asteroid traversed through a high perturbation region in the trailing edge of the Earth, likely contributing to the observed shift in the error curve in Fig. \ref{fig:errors_2005ED224}.
}

The geometry of approximation also helps explain a remarkable peculiarity concerning the planet Jupiter and deflected asteroids. Take, for example, Figures \ref{fig_defl_comp_2005ED224} and \ref{fig_defl_comp_3200Phaethon}, where the asteroids approach Jupiter within less than 10 SOIs. Nevertheless, the deflection predictions from the simplified models remain reasonably applicable. This behavior may seem strange, considering that Jupiter is the most massive body in the Solar System. Additionally, the asteroids would experience a greater ratio between Jupiter's gravitational perturbation and the Sun's gravitational attraction, being at such a close distance.

However, as in a gravity-assist, the momentum added by the perturbing body is more critical than a simple acceleration ratio. This is why Equation \ref{eq:dEdt} is essential to explain the phenomenon, as it incorporates the effect of the planet moving in its orbit~\cite{negri2017studying,Negri:2018:AnLiAp,negri2019lunar}.

\begin{figure}[!ht]
\centering
\subfigure[2005 ED224]{\includegraphics[width=.45\textwidth]{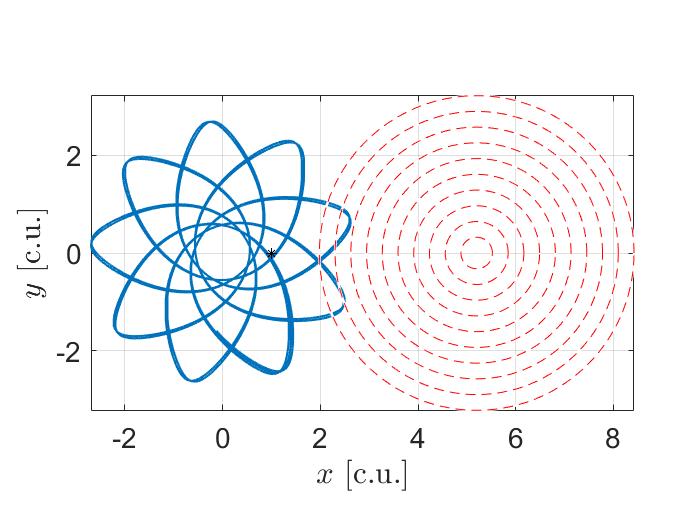}\label{fig_defl_closesyn_2005ED224}}
\subfigure[3200 Phaethon]{\includegraphics[width=.45\textwidth]{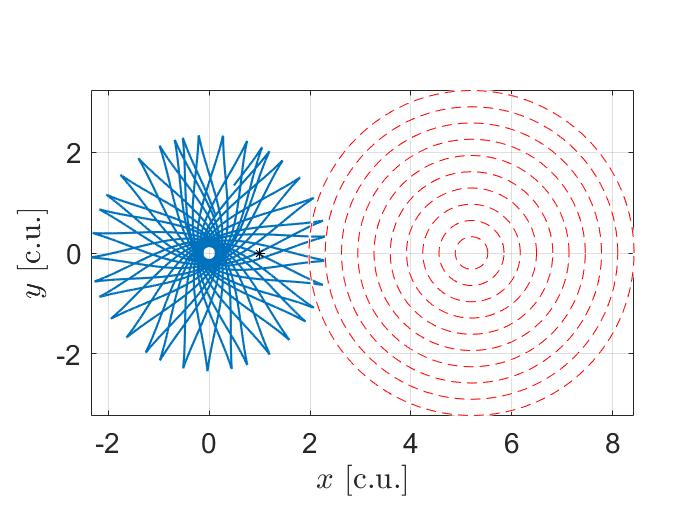}\label{fig_defl_closesyn_3200Phaethon}}

\caption{Projection of the trajectories of the asteroids 2005 ED224 and 3200 Phaethon, Sun and Jupiter fixed.}
\label{fig_defl_closesyn}
\end{figure}

Note in Figure \ref{fig_dEdt} that the energy variations are smaller as the proximity to $y=0$ increases. In fact, by simply inspecting Equation \ref{eq:dEdt}, it is evident that the energy variation is zero when $y=0$. This phenomenon explains the little influence of Jupiter in the case of the deflection of the selected asteroids. Since they do not cross Jupiter's orbit, the tendency is to approach the planet in regions where the energy change (and hence the other constants of motion as well) is close to zero. Figure \ref{fig_defl_closesyn} illustrates this fact by showing the projection of the trajectories of both asteroids in a synodic system where Jupiter is fixed. {Red dashed lines indicate distances from 1 to 10 SOIs from Jupiter.} As can be seen, these asteroids are confined to approach geometries that offer a small magnitude of $dE/dt$ since they do not cross Jupiter's orbit.

\begin{figure}[!ht]
\centering
\subfigure[Impulsive deflection]{\includegraphics[width=.45\textwidth]{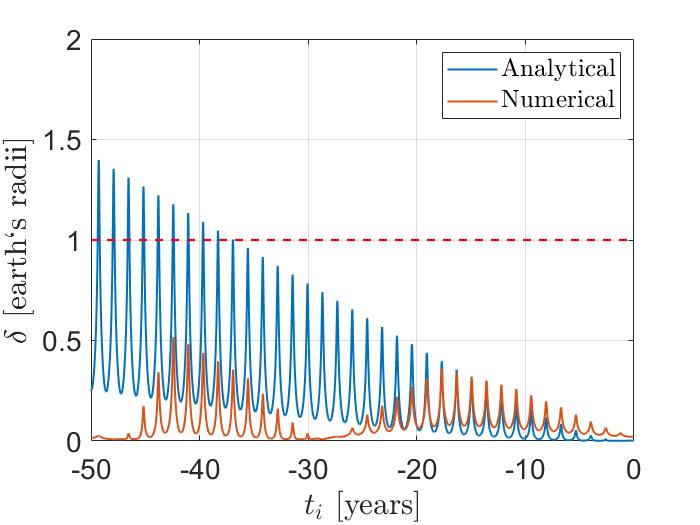}\label{fig_defl_impcomp_2008UB7_1mms}}
\subfigure[Low-thrust deflection]{\includegraphics[width=.45\textwidth]{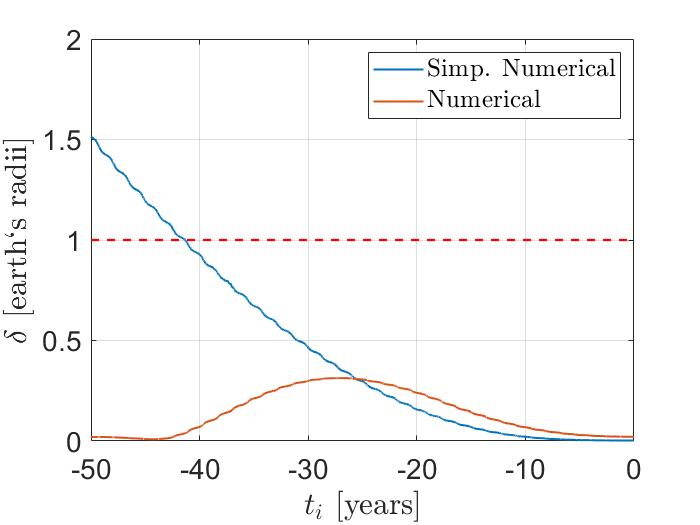}\label{fig_defl_lowcomp_2008UB7_1mms}} 
\caption{Deflection obtained with $\Delta V = 0.1$ cm/s or $A_t = 4\times 10^{-12}$ m/s$^2$ for the asteroid 2008 UB7.}
\label{fig_defl_comp_2008UB7_1mms}
\end{figure}

\begin{figure}[!ht]
\centering
\subfigure[Impulsive deflection]{\includegraphics[width=.45\textwidth]{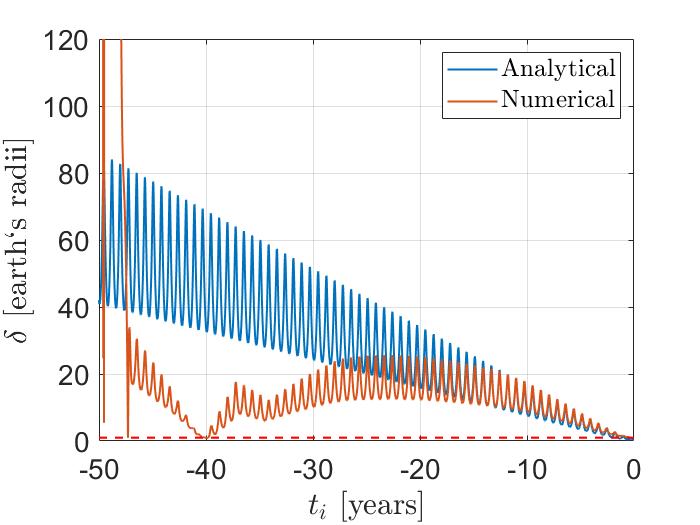}\label{fig_defl_impcomp_2020VW_10cms}}
\subfigure[Low-thrust deflection]{\includegraphics[width=.45\textwidth]{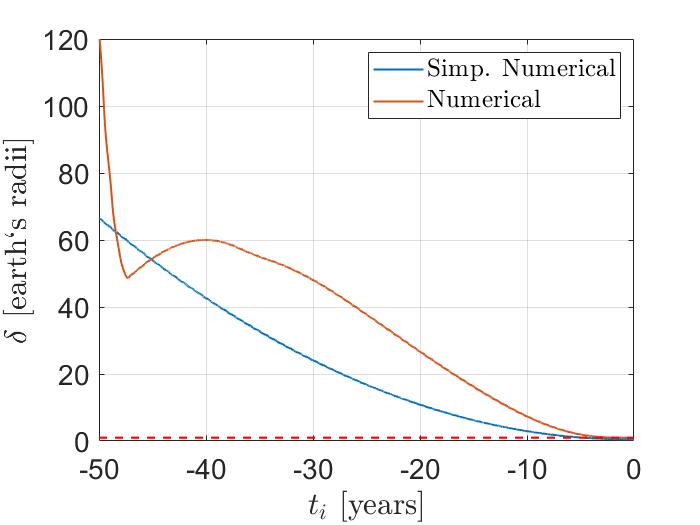}\label{fig_defl_lowcomp_2020VW_10cms}}

\caption{Deflection obtained with $\Delta V = 10$ cm/s or $A_t = 4\times 10^{-10}$ m/s$^2$ for the asteroid 2020 VW.}
\label{fig_defl_comp_2020VW_10cms}
\end{figure}

When reproducing Figures \ref{fig_defl_comp_2005ED224} to \ref{fig_defl_comp_3200Phaethon} for different magnitudes of impulse and thrust, with values reasonable to be applied, the conclusions remain unchanged. For example, two cases are reproduced in which the virtual asteroids are particularly disturbed by approaches to some of the planets. Figure \ref{fig_defl_comp_2008UB7_1mms} shows the deflections obtained considering $\Delta V = 0.1$ cm/s or $A_t = 4$ pm/s$^2$ (equivalent to a gravitational tractor of 10 tons at about 40 m) for the asteroid 2008 UB7. Figure \ref{fig_defl_comp_2020VW_10cms} represents the results for $\Delta V = 10$ cm/s or $A_t = 400$ pm/s$^2$ (equivalent to a gravitational tractor of 10 tons at about 410 m) for the asteroid 2020 VW.

{

{ We highlight the existence of analytical and semi-analytical methods in the literature for quantifying third-body effects~\cite{ross2007multiple,grover2009designing,masat2022flyby}. These methods represent promising initial approaches to address the impacts of shallow encounters. They may offer a viable solution that captures the effects of these encounters while maintaining simplicity for application in the preliminary design phase, particularly within optimization routines. We recommend further research efforts to delve into and expand upon these methods, rigorously showing their promising aspects and limitations for that purpose. For now, we propose a provisional guideline for mission analysts in the preliminary design of deflecting spacecraft trajectories (which will need a prediction of deflection). This guideline aims to assist until those studies are conducted.}

Specifically, we recommend that analysts consider the following criterion: if the asteroid \underline{crosses the orbit} of a planet and \underline{approaches} it within less than 15 times the radius of its sphere of influence at a time after the intercept time $t_i$, the analytical approximations tend to be invalid. Our findings indicate that any approach meeting these conditions resulted in the inability of the analytical model to predict the deflection. While some approaches greater than 15 SOIs, as exemplified by the case of the asteroid 2015 JJ, can lead to divergence, the majority of them had a limited impact on rendering the analytical models unsuitable. Conversely, all encounters within 15 SOIs rendered the analytical models unsuitable. 

{It is crucial to emphasize that these guidelines are derived from a limited dataset of asteroids presented in this study. Further research is necessary to unveil more intricate relationships. For example, the relative velocity in these shallow encounters is expected to play a crucial role in rendering analytical approximations unsuitable. With a larger sample size of asteroids, more intricate relationships can be identified and established as general guidelines or heuristic rules.}

For now, the importance of this study is in documenting and explaining the source of those discrepancies. The provisional guideline offers a practical means to assess the appropriateness of the best simplified methods currently available for the preliminary design phase of a real case scenario. While it is conceivable to already devise a ``complex approximation'' to estimate the perturbation added by the shallow encounter with the planet, such an approach would inevitably involve obtaining the entire trajectory of the deflected asteroid, examining its displacement from the original trajectory, considering its minimal change in geometry during the encounter, and assessing the net effects of the perturbing body on the deflected asteroid. However, this ``complex approximation'' would not be practically useful in optimization routines during the preliminary design phase of a trajectory for a deflecting spacecraft, when analytical and simplified models are preferable. It would be much easier and practical to go straight for an accurate numerical solution in this case.

In this context, the proposed guideline proves more practical as it allows for the straightforward exploration of a wide design space. It facilitates the discrimination between conditions in which the designer can have greater confidence that the trajectory obtained with the analytical approximations will remain feasible during the later refinement phase and those that require extra care and attention. The later may potentially necessitate the direct application of more costly numerical methods, impacting both the time spent in the preliminary design phase and the extent of the design space that can be explored.
}

{

\subsection{Impact of the Findings on Trajectory Design}

Suppose a mission analyst is tasked with designing the preliminary trajectory for a deflecting spacecraft through kinetic impact. Gravity-assists, as demonstrated by Vasile and Colombo \cite{vasile2008optimal} and Negri \cite[Chp. 4]{negri2022tese}, can increase the chance of success by inexpensively boosting the orbital energy of the spacecraft and producing more favorable conditions for colliding the spacecraft with the asteroid. For this case, predicting the required deflection is crucial in solving the multiple gravity-assist problem to find trajectories leading to the desired velocity change, as exemplified by Negri \cite[Chp. 4]{negri2022tese}.

In the preliminary design of gravity-assist trajectories, analytical approximations are often employed to explore a wide design space within a reasonable time frame. Similarly, for a deflecting spacecraft, a simple analytical prediction of deflection, such as the one presented in Section \ref{sec:vasile}, would be preferable.

Our study reveals that even shallow encounters, occurring a dozen Spheres of Influence away, can render the best analytical approximations unsuitable. The critical question then becomes how a designer can anticipate whether the trajectories found in the preliminary design phase, e.g., in the case of multiple gravity-assists, will yield accurate estimates in later design phases.

Without our study, the designer might invest significant time exploring the design space only to find that some preliminary trajectories fail to produce near-optimal solutions in later design phases, even without entering the Sphere of Influence (SOI) of any planet or having resonant returns within the Earth's SOI. Our study provides valuable insights into what designers should expect when looking for such trajectories.

If the asteroid does not cross the orbit of any planet, the designer can confidently apply analytical models, as they are less likely to fail due to planetary perturbations. However, if the asteroid crosses a planet's orbit and experiences a shallow encounter, the designer should expect that analytical estimates will generally fail. The confidence level decreases especially if the encounter is within 15 SOIs, and in such cases, the designer might opt for direct numerical predictions of deflections without investing time in generating a large set of preliminary trajectories with the analytical models, as they are likely to be inaccurate.

Again, this study represents only a step in the exploration of this phenomenon. Subsequent research endeavors should build upon these findings, providing clearer guidelines on how to address these effects. Defining precise thresholds and criteria for when to choose between the analytical models presented in this study and other potentially proposed models is crucial. As our understanding deepens, a more comprehensive framework for choosing the most appropriate model for specific mission scenarios will emerge. The contributions of each planet are also worth further investigation. For instance, should we expect a similar behavior if the asteroid crosses the orbit of Mercury or Mars at 15 SOIs as well? We unfortunately could not capture any shallow encounter with Mars or Mercury in our simulations, despite our efforts in trying to select an asteroid with this requirement.

An intriguing avenue for trajectory design exploration lies in utilizing these shallow encounters as gravity-assists to enhance deflection, as suggested by Chagas et al. \cite{chagas2022deflecting} (albeit not specifically treating shallow encounters). Leveraging the phenomenon revealed in this study, a potentially smaller spacecraft, in the case of kinetic impact or gravity-tractor, could induce a considerably larger deflection. However, comprehensive studies must precede the credence of such a proposition. Contingencies during or after the deflection, while potentially enhancing deflection, may also escalate the probability of asteroid collision with Earth, particularly if the deflection was proposed due to Earth being on the fringes of the uncertainty ellipsoid, or worsen collision conditions, such as increased collision velocity.

Considering shallow encounters in tandem with kinetic impact introduces potential challenges, even if shallow encounters are not intended to boost deflection. Such encounters may pose difficulties in deflection options lacking active control for correcting the deflected asteroid's orbit post-shallow encounter. Preemptive measures, such as having backup deflecting spacecraft ready, become essential in such scenarios to address contingencies. Similarly, ``passive control'' deflections, like sublimating material from the asteroid's surface, may prove problematic in shallow encounter scenarios. Active control approaches, such as using gravity tractors, offer a more favorable solution in addressing such scenarios, providing the spacecraft with the capability to correct the deflected asteroid's orbit post-shallow encounter, which also makes the use of shallow encounters as gravity-assists more favorable. In summary, further studies should be conducted to understand the risks posed by those shallow encounters in different deflection approaches, as well as make use of them to boost deflection.

{

In designing shallow gravity-assists to enhance deflection, our general guidelines offer limited assistance. They can only inform the designer that such a possibility exists for the specific asteroid under analysis. Utilizing well-established design tools for spacecraft gravity-assists under analogous circumstances, such as the Tisserand-Poincaré graph~\cite{campagnola2010endgame} or Keplerian map~\cite{ross2007multiple}, holds promise for the preliminary design of such maneuvers. However, these tools may require some modifications to account for deflection prediction with simplicity and robustness.
}
}

\section{Conclusion}

{
This study delved into investigating the influence of other celestial bodies on the deflection of asteroids, employing both analytical and numerical models. The analysis revealed that the impact of perturbing bodies on deflection is more extensive and intricate than previously assumed in the literature. While the best analytical and semi-analytical estimates can provide reasonably accurate results for some cases, there are scenarios where significant discrepancies arise.

A crucial finding was the influence of an asteroid's approach to perturbing bodies during or after deflection. Notably, even distant shallow encounters with planets can perturb the asteroid's trajectory to the same order of magnitude as the deflection itself, rendering simple analytical models invalid in such cases. The geometry of approximation plays a crucial role, with specific encounter geometries leading to a considerable divergence between simplified and numerical models. {The same can be expected for other intricacies of the dynamics, such as the relative velocity, which is almost tacit knowledge among designers of outer planetary system multiple gravity-assists.}

To aid in the preliminary design of deflection missions, we give provisional general guidelines. The simplified approximation for deflection tends to be invalid if the asteroid crosses the orbit of a planet and approaches it within less than 15 times the radius of its sphere of influence after the applied deflection. Any approach meeting these conditions resulted in the inability of the analytical model to predict the deflection. While some approaches greater than 15 SOIs can lead to divergence, most of them had a limited impact on rendering the analytical models unsuitable.

In the context of trajectory design for a deflecting spacecraft, especially in scenarios involving multiple gravity-assists, where predicting the required deflection is essential, our study highlights the need for careful consideration. If the asteroid does not cross the orbit of any planet, analytical models can be confidently applied to predict the deflection. However, if the asteroid crosses a planet's orbit and experiences a shallow encounter, the confidence in analytical estimates decreases, especially if the encounter is within 15 SOIs. In such cases, opting for numerical solutions directly, without investing time in generating a large set of preliminary trajectories, is a more prudent approach. {However, the proposed guideline is of little use if the aim is to make use of the shallow encounter to boost the deflection, only informing the designer when such an opportunity is available.}}

\appendix

\section*{Tables}
\label{app:tables}

\setlongtables
\begin{landscape}
\begin{longtable}{llllllll}
\caption{\label{tab:osc_horizons}Osculating orbital elements for the considered asteroids.} \\
\hline
Asteroid & $a$ [AU] & $e$ & $i$ [$^\circ$] & $\Omega$ [$^\circ$] & $\omega$ [$^\circ$] & $M$ [$^\circ$] & $JD$ \\
\hline
\endfirsthead
\endhead
\hline
\endlastfoot
2005 ED224 & 1.9104727634 & 0.6599137769 & 31.9242807796 & 170.1460792025 & 277.5419453853 & 300.9979911173 & 2459986.5 \\
2008 EX5 & 1.3637869020 & 0.3921667357 & 3.3762480685 & 15.6516027929 & 66.5603750333 & 251.2376806053 & 2478094.5 \\
2008 UB7 & 1.2358466255 & 0.5938144619 & 2.0465323634 & 218.7381945812 & 288.4636746512 & 292.7155382079 & 2474828.5 \\
2009 JF1 & 1.8927712462 & 0.7382924981 & 6.1505990860 & 45.5247150416 & 281.4107281902 & 300.2059121420 & 2459600.5 \\
2010 RF12 & 1.0555593928 & 0.1871129963 & 0.9114680037 & 162.5758249980 & 266.8796828793 & 265.0179362689 & 2486459.5 \\
2015 JJ & 1.0429924820 & 0.1498084466 & 18.9941064156 & 222.6682531868 & 259.2330847116 & 250.8953368455 & 2492365.5 \\
2020 VW & 0.8396131472 & 0.3477983731 & 3.0624690103 & 219.6865079915 & 42.6627738683 & 63.9039653057 & 2478847.5 \\
2021 EU & 2.0773609642 & 0.7216949667 & 3.8570482564 & 155.9490078152 & 270.8098365840 & 321.8086041717 & 2472208.5 \\
3200 Phaethon & 1.2716436178 & 0.8890253987 & 22.7109907367 & 264.1758704361 & 323.1898702235 & 313.4367740049 & 2485830.5 \\
\end{longtable}
\vspace{-8mm}
\end{landscape}

\setlongtables
\begin{landscape}
\begin{longtable}{llllllll}
\caption{\label{tab:osc_calculados}Osculating orbital elements for the considered asteroids after optimizations ensuring impact in simulations using CRNBP.} \\
\hline
Asteroid & $a$ [AU] & $e$ & $i$ [$^\circ$] & $\Omega$ [$^\circ$] & $\omega$ [$^\circ$] & $M$ [$^\circ$] \\
\hline
\endfirsthead
\endhead
\hline
\endlastfoot
2005 ED224 & 1.9110955078 & 0.6592630085 & 31.9239330982 & 170.1733183595 & 277.0027823224 & 7.0202856620 \\
2008 EX5 & 1.3576571615 & 0.3949941837 & 3.3847205353 & 17.3728683050 & 68.3544423264 & 309.9589534417 \\
2008 UB7 & 1.2354871130 & 0.5935261954 & 2.0470053214 & 219.7194257208 & 289.6693134514 & 298.3863051886 \\
2009 JF1 & 1.8924595146 & 0.7382974846 & 6.1510316994 & 45.4993775387 & 280.8691712217 & 302.3546096772 \\
2010 RF12 & 1.0510291265 & 0.1890062129 & 0.8549225818 & 162.1814477568 & 266.2796834888 & 265.9572348632 \\
2015 JJ & 1.0395704538 & 0.1494259445 & 18.9951848815 & 224.2253334875 & 263.7557363965 & 263.9005040212 \\
2020 VW & 0.8402436619 & 0.3481420812 & 3.0621561884 & 219.7309166538 & 41.3536702061 & 65.5589395093 \\
2021 EU & 2.0770434581 & 0.7214227546 & 3.8541725742 & 155.7460268300 & 270.3184408436 & 334.4393695649 \\
3200 Phaethon & 1.2717424008 & 0.8886229663 & 22.7127670180 & 261.8743689654 & 325.5200455601 & 312.3997120626 \\
\end{longtable}
\vspace{-8mm}
\end{landscape}


\begin{table}[!ht]
    \centering
    \caption{\label{tab:dist_enc}shallow encounters below 20 SOIs between each asteroid and the planets.}
    \footnotesize
    \begin{tabular}{l|p{1.25cm}|S[round-mode=places, round-precision=1, table-format=3.1]|S[round-mode=places, round-precision=1, table-format=2.1]|S[round-mode=places, round-precision=1]}
        Asteroid & Encounter Planet & \multicolumn{1}{p{1.4cm}|}{{$t$ [years]}} & \multicolumn{1}{p{1.4cm}|}{{Closest Approach [SOIs]}} & \multicolumn{1}{c}{{$\alpha$ [$^\circ$]}} \\
        \hline
        \hline
2005 ED224 & Earth & -13.481233 & 16.324871 & 82.609613 \\ 

2005 ED224 & Earth & -36.985908 & 16.289337 & -10.728764 \\ 
    \hline
2008 EX5 & Earth & -7.644157 & 14.116098 & 35.756478 \\ 

2008 EX5 & Earth & -18.956549 & 16.866998 & -35.132851 \\ 

2008 EX5 & Earth & -26.610306 & 8.424013 & -50.750600 \\ 

2008 EX5 & Earth & -49.013834 & 10.078697 & 4.353843 \\ 
\hline
2008 UB7 & Earth & -10.989198 & 15.165756 & -6.743491 \\ 

2008 UB7 & Earth & -35.394819 & 5.737669 & 39.802530 \\ 

2008 UB7 & Earth & -46.391177 & 3.766679 & -74.051007 \\ 

2008 UB7 & Venus & -17.618057 & 10.374155 & 58.155473 \\ 

2008 UB7 & Venus & -24.645523 & 8.019833 & 76.765781 \\ 

2008 UB7 & Venus & -35.467850 & 19.550195 & 6.130313 \\ 

2008 UB7 & Venus & -42.495113 & 5.728824 & -38.032527 \\ 
\hline
2009 JF1 & Earth & -13.016490 & 18.580610 & 11.139926 \\ 

2009 JF1 & Venus & -2.548469 & 15.955356 & -42.213835 \\ 

2009 JF1 & Venus & -33.655467 & 2.150715 & -78.422453 \\ 
\hline
2010 RF12 & Earth & -12.642220 & 19.306271 & 62.495508 \\ 

2010 RF12 & Earth & -14.009254 & 10.409798 & 1.263399 \\ 

2010 RF12 & Earth & -39.523917 & 1.169981 & 16.990099 \\ 

2010 RF12 & Earth & -40.993887 & 6.228697 & 1.458728 \\ 
\hline
2015 JJ & Earth & -18.011941 & 18.509893 & 8.938634 \\ 

2015 JJ & Earth & -33.504522 & 7.999139 & -40.209590 \\ 

2015 JJ & Earth & -34.987988 & 18.400758 & -6.575829 \\ 
\hline
2020 VW & Earth & -2.773561 & 9.315333 & -70.244188 \\ 

2020 VW & Earth & -9.972483 & 13.613794 & 49.175439 \\ 

2020 VW & Earth & -12.768097 & 15.892974 & 38.763661 \\ 

2020 VW & Earth & -27.000197 & 17.492270 & -21.101503 \\ 

2020 VW & Earth & -29.817501 & 18.303991 & -64.804684 \\ 

2020 VW & Earth & -37.001058 & 4.187286 & -25.858816 \\ 

2020 VW & Earth & -39.792213 & 13.825850 & -57.208189 \\ 

2020 VW & Earth & -46.999802 & 0.892923 & 32.068512 \\ 

2020 VW & Earth & -49.765815 & 16.954693 & 35.792630 \\ 

2020 VW & Venus & -0.129053 & 19.886724 & 22.381736 \\ 

2020 VW & Venus & -31.161136 & 10.829831 & 36.492036 \\ 

2020 VW & Venus & -34.243032 & 6.720931 & 78.476765 \\ 

2020 VW & Venus & -37.324776 & 9.192789 & -80.853144 \\ 

2020 VW & Venus & -40.402087 & 10.928690 & 80.149823 \\ 

2020 VW & Venus & -43.480139 & 15.234849 & 54.818608 \\ 
\hline
2021 EU & Earth & -2.993720 & 8.591710 & -4.031326 \\ 

2021 EU & Earth & -32.495393 & 1.102062 & 1.800032 \\ 

2021 EU & Earth & -35.485877 & 8.636404 & -3.866913 \\ 

2021 EU & Earth & -38.479026 & 14.747137 & -6.712090 \\ 

2021 EU & Venus & -8.765841 & 8.281019 & -86.688093 \\ 
\hline
3200 Phaethon & Earth & -32.990137 & 17.449544 & -27.828659 \\ 

3200 Phaethon & Earth & -43.009849 & 14.224499 & 30.465679 \\

        \hline
    \end{tabular}
\end{table}
\normalsize

\section*{Funding Sources}

The authors wish to express their appreciation for the support provided through the grant $\#$ 301338/2016-7 from the  National Council for Scientific and Technological Development (CNPq); grants $\#$ 2021/10853-7, 2017/20794-2, 2016/24561-0 from S\~ao Paulo Research Foundation (FAPESP); and the financial support from the Coordination for the Improvement of Higher Education Personnel (CAPES).

\bibliographystyle{ieeetr}
\bibliography{referencia}

\end{document}